\documentstyle[11pt,aaspp4,psfig]{article}

\newcommand{\Mo}{\mbox{$\rm M_\odot$}}
\newcommand{\Lo}{\mbox{$\rm L_\odot$}}

\newcommand{\ivol}{\mbox{$\rm cm^{-3}$}}

\newcommand{\isec}{\mbox{s$^{-1}$}}

\newcommand{\Ne}{\mbox{$N_{\rm e}$}}
\newcommand{\ten}[1]{\mbox{$10^{#1}$}}
\newcommand{\xten}[1]{\mbox{$\times 10^{#1}$}}
\newcommand{\wl}{\mbox{$\lambda$}}
\newcommand{\forb}[2]{\mbox{$[{\rm #1\, #2}]$}}
\newcommand{\ha}{\mbox{H$\alpha$}}
\newcommand{\hb}{\mbox{H$\beta$}}
\newcommand{\oiii}{\forb{O}{III}}
\newcommand{\oii}{\forb{O}{II}}
\newcommand{\oi}{\forb{O}{I}}
\newcommand{\nii}{\forb{N}{II}}
\newcommand{\sii}{\forb{S}{II}}

\begin{document}

\lefthead{Macchetto et al.} 
\righthead{Super-massive Black-Hole of M87}

\title{The supermassive black hole of M87 and the kinematics 
of its associated gaseous disk
\footnote{Based on
observations with the NASA/ESA Hubble Space Telescope, obtained at the
Space Telescope Science Institute, which is operated by AURA, Inc.,
under NASA contract NAS 5-26555 and by STScI grant GO-3594.01-91A}}

\author{F. Macchetto\altaffilmark{2}, A. Marconi\altaffilmark{3},
D.J. Axon\altaffilmark{2,4}}
\affil{Space Telescope Science Institute \\
       3700 San Martin Drive\\
       Baltimore, MD 21218}
\author{A. Capetti}
\affil{Scuola Superiore di Studi Superiori Avanzati\\
Via Beirut 2-4\\
34014 Trieste, Italy}
\author{W. Sparks}
\affil{Space Telescope Science Institute}
\author{P. Crane}
\affil{European Southern Observatory\\
Karl-Schwarzschild-Str. 2\\
D-85748 Garching bei M$\ddot{\rm u}$nchen, Germany}

\altaffiltext{2}{Affiliated to the Astrophysics Division, Space Science
Department, ESA}
\altaffiltext{3}{Dipartimento di Astronomia e Scienza dello Spazio,
Universit\`a di Firenze,\\ Largo E. Fermi 5, I--50125, Italy}
\altaffiltext{4}{On leave from the University of Manchester}

\authoremail{macchetto@stsci.edu}

\begin{abstract}
We have obtained long-slit observations of the circumnuclear region
of M87 at three different locations, with a spatial sampling of 0\farcs028
using the Faint Object Camera f/48 spectrograph on board HST.
These data allow us to determine the rotation  curve of the 
inner $\sim$ 1\arcsec\ of the ionized gas disk  in \oii\wl 3727 
to a distance as close as 0\farcs07 ($\simeq 5$pc) to the dynamic 
center, thereby significantly improving on both the spatial resolution
and coverage of previous FOS observations. We have modeled the kinematics 
of the gas under the assumption of the existence of both a 
central  black hole and an extended  central mass distribution, 
 taking into account the effects of the instrumental PSF, the 
intrinsic luminosity distribution of the line, and the finite size of
the slit.
We find that the central mass must be concentrated within a sphere
whose maximum radius is 0\farcs05 ($\simeq$3.5pc) and show that both
the observed rotation curve and line profiles are consistent with a thin--disk
in keplerian motion. We conclude that the most likely explanation for
the observed motions is the presence of a supermassive black hole and
derive a value of $M_{BH} = (3.2\pm 0.9)\xten{9}$\Mo\ for its mass.
\end{abstract} 


\section{Introduction}

The presence of massive black holes at the center of galaxies is widely
believed to be the common origin of the phenomena associated with so-called
Active Galactic Nuclei. The black hole model is very appealing
because it provides an efficient mechanism that converts 
gravitational energy, via
accretion, into radiation within a very small volume as required by the
rapid variability of the large energy output of AGN
(e.g. \cite{blandford}).

The standard model comprises a central black hole with mass in the range
$\simeq\ten{6}-\ten{9}\Mo$ surrounded by an
accretion disk which releases gravitational energy. The
radiation is emitted thermally at the local black--body temperature and is
identified with the ``blue--bump'', which accounts for the majority of the
bolometric luminosity in the AGNs. The disk possesses an active corona,
where infrared synchrotron radiation is emitted along with thermal
bremsstrahlung X-rays. The host galaxy supplies this disk with gas at
a rate that reflects its star formation history and, possibly, its
overall mass (\cite{magorrian}) thereby accounting 
for the observed luminosity evolution. Broad emission lines
originate homogeneously in small gas clouds of density $\simeq\ten{9}\ivol$
and size $\simeq$1 AU in random virial orbits about the central
continuum source. Plasma jets are emitted perpendicular to the disk.
At large radii, the material forms an obscuring torus of cold molecular
gas. Orientation effects of this torus to the line-of-sight 
naturally account for the differences between some
of the different classes of AGN (see \cite{antonucci} for a
review). While this broad picture has been supported and refined by a number 
of observations, direct evidence for the existence of accretion disks
around supermassive black holes is sparse (see, however, \cite{livio})
and detailed measurements of their physical
characteristics are conspicuous by their absence.

The giant elliptical galaxy M87 at the center of the Virgo cluster, at a
red-shift of 0.0043 (\cite{z}),
is well known for its spectacular, apparently one--sided jet, 
and has been studied extensively across the electromagnetic spectrum.
Ground--based observations first revealed the presence of a cusp--like region
in its radial light profile accompanied by a rapid rise in the stellar
velocity dispersion and led to the suggestion that it contained a massive
black hole (\cite{young}, \cite{sargent}).
Stellar dynamical models of elliptical galaxies showed however
that these velocity dispersion rises did not necessarily imply the
presence of a black hole, but could instead be a consequence of an
anisotropic velocity dispersion tensor in the central 100pc of a triaxial elliptical potential
(e.g. \cite{duncan}, \cite{mamon}).
Considerable controversy has surrounded this and numerous other attempts
to verify the existence of the black hole in M87 and other
nearby giant ellipticals using ground based stellar dynamical studies
(e.g. \cite{dressler}, \cite{vandermarel}). 
To-date the best available data remains ambiguous largely because of the difficulty
of detecting the high-velocity wings on the absorption lines which are the hallmark
of the black hole. 

One of the major goals of HST has been to establish or refute the
existence of  black holes in  active galaxies by probing the dynamics
of AGN at much smaller radii than can be achieved from the ground.

HST emission line imagery  (\cite{crane}, \cite{wfpc94}) of  M87 has
lead to the discovery of a small scale disk of ionized gas surrounding its nucleus
which is oriented approximately perpendicularly to the synchrotron jet.
This disk is also  observed in both  the optical and UV continuum
 (\cite{macchettoA} and 1996b).
Similar gaseous disks have also been found in the nuclei of a  number of 
other massive galaxies (\cite{ferrarese}, \cite{jaffe}). 

Because of surface brightness limitations on stellar dynamical studies at 
HST resolutions, the kinematics of such disks are in practice likely to be 
the only way to determine if a central black hole exists in all but the 
very nearest galaxies.  In the case of M87 FOS observations at two 
locations on opposite sides of the nucleus separated by 0\farcs5 showed a 
velocity difference of $\simeq$1000 km \isec, a clear indication of rapid 
motions close to the nucleus (\cite{fos94}). By {\it assuming} that the gas 
kinematics determined at these and two additional locations arise in a thin 
rotating keplerian disk, \cite{fos96} estimated the central mass of M87 
is $\simeq$2\xten{9}\Mo\ with a range of variation between 1 and 
3.5\xten{9}\Mo.  Currently this result provides the most convincing 
observational evidence in favour of the black hole model.  Implicit in 
this measurement of the mass of the central object, however, is the 
assumption that the gas motions in the innermost regions reflect keplerian 
rotation and not the effects of non gravitational forces such as 
interactions with the jet.  Establishing the detailed kinematics of the disk 
is therefore vital.

In this paper we present new FOC,f/48 long-slit spectroscopic observations
of the ionized circumnuclear disk of M87 with a pixel size of 0\farcs028.
They provide us with a  \oii\wl 3727 rotation curve in three different
locations on the disk and which extend up to a distance of $\sim$ 1\arcsec. 
We show that the observations are consistent with a thin--disk in keplerian
motion, which explains the observed rotation curve and line profiles, and
we derive a mass of $M_{BH} = (3.2\pm 0.9)\xten{9}$\Mo\ within
a radius of $\simeq$5pc for the central black hole.

The plan of the paper is as follows: the observations and data reduction are
described in sections 2 and 3. In section 4 we use the
current data and previous HST images
to constrain the precise location of the slit with respect to the nucleus 
of M87.
The results of the observations are presented in section 5 and compared to 
HST/FOS observations in section 6. In section 7 and 8 we discuss
the fitting procedures to the observed rotation curve and line profiles 
with increasingly sophisticated models and in section 9 we compare
the observed line profiles with the predictions from the keplerian
model of a thin disk.
Limits on the distributed mass are discussed in section 10 and 
the conclusions are given in section 11.

Following \cite{fos94} we adopt a distance to M87
of 15 Mpc throughout this paper, whence 0\farcs1 correspond to 7.3 pc.

\section{Observations}

The circumnuclear disk of M87 was observed using the Faint Object Camera
f/48 long--slit spectrograph on board the Hubble Space Telescope
on July 25$^{\rm th}$, 1996 at resolutions of  1.78\AA\ 
and 0\farcs0287 per pixel along the dispersion and slit directions,
respectively.
The F305LP filter was used to isolate the first order
spectrum which covers the 3650--5470 \AA\ region and therefore includes
the \oii\wl 3727, \hb\wl 4861 and \oiii\wl\wl 4959, 5007 \AA\  emission lines.
An interactive acquisition (integration time 297s)
1024x512 zoomed image was obtained with the f/48 camera
through the F140W filter to accurately locate
the nucleus.
The slit, 0\farcs063x13\farcs5 in size, was
positioned on the gas disk at a position angle of 47.3$^\circ$ and
initially spectra with exposure times
of 2169 seconds were taken in the 1024x512 non--zoomed mode
at 3 locations separated by 0\farcs2 centered approximately on the
nucleus.
These observations were used to derive the actual location
of the nucleus and HST was then repositioned using a small angle
maneuver to this location (which actually turned out to be virtually coincident with the
position of the second of the three scans).
This allowed us to position the slit to within 0\farcs1 of the nucleus
(see section \ref{sec:impact}).
A further higher  signal-to-noise spectrum at this location was then
obtained with a total exposure time of 7791 seconds, built from 3 shorter
exposures, 2597 seconds in duration. The accuracy of the HST
small angle maneuvers is known to be a few milliarcseconds
and this is in agreement with 
the close correspondence between the four spectra taken on the nucleus 
during the spatial scan. 

\begin{figure}
\centerline{\psfig{file=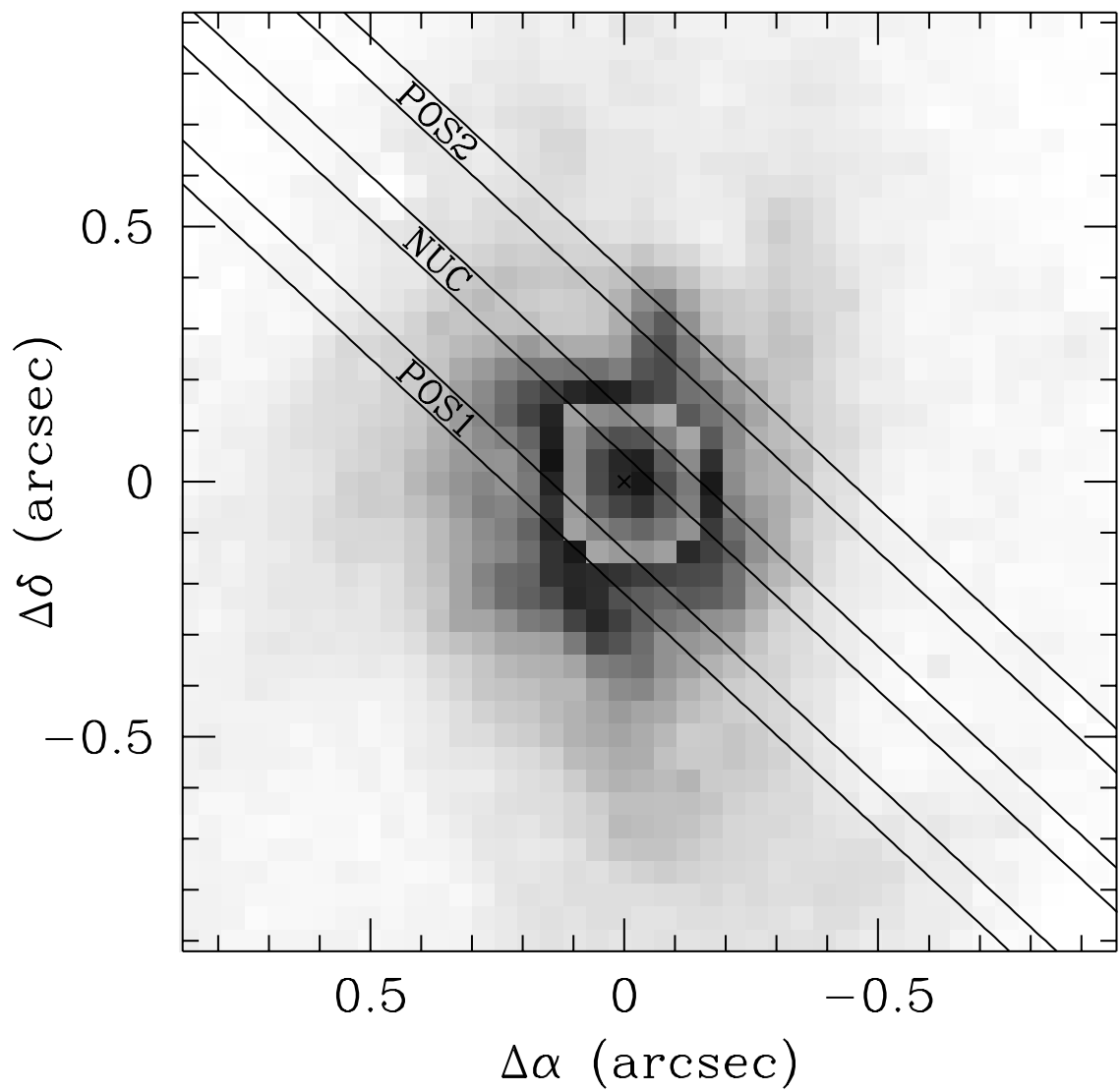,height=15cm,angle=-90}}
\caption{\label{fig:slitpos} Positions of the slit during
the observations compared with the \ha+\nii\ image of the M87 disk
from the WFPC2 archive. The gray levels are between 0 and 40\%
of the nuclear peak in the outer region. The nucleus has been 
rescaled to be displayed within this range of values.
North is up and east is to the left.}
\end{figure}

The actual slit positions,
as derived in section \ref{sec:impact}, are displayed in Fig. \ref{fig:slitpos}
superposed on the \ha+\nii\ image published by \cite{wfpc94} which
we retrieved from the HST archive.
Hereafter we will refer to them as POS1, NUC and POS2
from South--East to North--West, respectively.
In Tab. 1 we list the datasets obtained for M87 and those
which were used for the calibration. 

\section{\label{sec:calib} Data reduction}

The raw FOC data suffer from geometric 
distortion, i.e., the spatial relations between objects on the sky are not
preserved in the raw images produced by the FOC cameras.
This geometric distortion can be viewed as originating from two
distinct sources. The first of these, optical distortion, is external
to the detectors and arises because of the off--axis nature of the instrument
aperture. The second, and more significant source of distortion
is the detector itself, which is magnetically focused.

All frames, including those used for calibration, were geometrically
corrected using the equally spaced grid
of reseau marks which is etched onto the first of the bi--alkali photocatodes
in the intensifier tube (\cite{focman}). 
This geometric correction takes into account both the
external and internal distortions.
The positions of the 
reseau marks were measured on each of the internal flat--field frames
which were interspersed between the spectra.
The transformation between these positions and an equally spaced 9x17 artificial grid 
was then computed by fitting bi-dimensional Chebyshev polynomials
of 6$^{\rm th}$ order in x and y terms and 5$^{\rm th}$ order
in the cross terms.
This transformation was applied to the science images
resulting in a mean uncertainty in the reseau position
of 0.10--0.25 pixels, depending on the signal-to-noise (SNR) of the flat frames.

In addition, geometric distortion is also induced on the slit and dispersion
directions by the spectrographic mirror and the grating.
The distortion in the dispersion direction was determined by
tracing the spectra of two stars taken in the core of the 47 Tucanae
globular cluster.
These stars are $\simeq$ 130 pixels apart and almost
at the  opposite extremes of the slit.
The distortion along the slit direction was determined by tracing the
brightness distribution along the slit of the planetary nebula
NGC 6543 emission lines. 
Ground--based observations (Perez, Cuesta, Axon and Robinson, in preparation) 
indicate that the distortion induced by the velocity field of
the planetary nebula are negligible (less than 0.5 \AA) at the
f/48 resolution. After rectification the spectra of the 47 Tuc
stars and that of NGC 6543 were
straight to better than 0.2 pixels on both axes.
The wavelength calibration was determined from the geometrically corrected
NGC 6543 spectrum. The reference wavelengths were again derived from
the ground--based observations.  The residual uncertainty on the wavelength
calibration is less than 0.2\AA.
As a result of these procedures we obtained images with the
dispersion direction along columns and the slit direction along rows.
The pixel sizes are 0\farcs0287 in the spatial
direction and 1.78\AA\ along the dispersion direction.

The relatively small width of the lines of NGC 6543 (FWHM$<$ 100 km \isec)
allows us to estimate the instrumental broadening to be
$\simeq$ 430$\pm$30 km \isec.
From the luminosity profile of the 47 Tuc stellar spectra the instrumental
broadening along the spatial direction is 0\farcs08$\pm$0\farcs02.

The contributions to the total error budget from the various calibration
steps can be summarized as follows:
\begin{itemize}
\item[i)] geometric correction with the reseau marks has a residual error
of 0.10--0.25 pixels;
\item[ii)] rectification of the dispersion direction has a residual error
less than 0.2 pixels; 
\item[iii)] rectification along the slit direction has a residual error of 
0.2 pixels;
\item[iv)] the error due to the intrinsic distortions of the planetary
nebula velocity field is less than 0.5\AA, i.e. 0.3 pixels;
\item[v)] the residual error in the wavelength calibration is less
than 0.15\AA.
\end{itemize}

Combining all the uncertainty terms in quadrature we estimate a maximum
uncertainty of 0.45 pixels (0.8 \AA\, which correspond to 65 km \isec\ at 3727 \AA)
and 0.28 pixels (8 mas) in the dispersion and spatial directions respectively.
Moreover, when we restrict ourselves to a small region of the detector, 
corresponding to a single spectral line, both uncertainties are negligible 
compared to those arising from the SNR of the data.

The distortion correction and wavelength calibration were applied
to the geometrically corrected M87 spectra and, 
as a check on our error budget, we traced the nuclear
continuum emission. We found that the continuum was straight to better than 
1 pixels across the whole spectral range and 0.5 pixels if we excluded
the low SNR region of  the spectrum redward of 4800 \AA\ which was not
used in our  analysis.

The imperfect repositioning of the spectrographic mirror, which moves 
between flat--field and source exposures, caused shifts between successive 
spectra in both the spatial and spectral directions.  By comparing the 
internal consistency of the four independent spectra of the nucleus of M87 
we have determined that these shifts range from 1 to 4 pixels in the 
spatial direction and are less than 2 pixels in the dispersion direction.  
The four spectra were aligned to an accuracy of 0.02 pixels by 
cross-correlating the flux distribution of the \oii\ line and 
co--added.  The relative zero points of the other slit positions cannot be 
determined because the continuum is too weak to be detected.  In the 
following analysis we will therefore conservatively allow for zero--point 
shifts in both the spatial and velocity directions of up to 4 pixels.

The background emission was subtracted in all the spectra by means of a
spline fit along the slit direction after masking out the regions where
line or continuum emission is detected.  Similarly the continuum under
the lines was subtracted with a first order polynomial fit along the
dispersion direction.

\section{\label{sec:impact} Determination of the slit location}

Given the 0\farcs2 step of the spatial scan and the slit width of 
0\farcs063, the ``impact parameter'', $b$, the minimum distance between the 
center of the slit and the nucleus is  constrained to be smaller than 
0\farcs1 by our  observing strategy . We accurately determine $b$ by 
comparing the flux measured from each of our three slit positions with the 
brightness profile derived from a previous FOC,f/96 HST image in the F342W. 
This filter covers a similar spectral range and includes the dominant 
(Sec.\ref{sec:results}) line in our spectra, \oii.
The F342W filter has a width of about 670 \AA\ and 
the scale of the f/48 spectra along
the dispersion direction is 1.78 \AA/pixels. To correctly synthesize
the spectral energy distribution transmitted by the F342W filter,
for each slit we collapsed 376 spectral channels
around the \oii\ line and then measured the flux by
co--adding 30 pixels (0\farcs9)
around the peak in the slit direction.

\begin{figure}
\centerline{\psfig{file=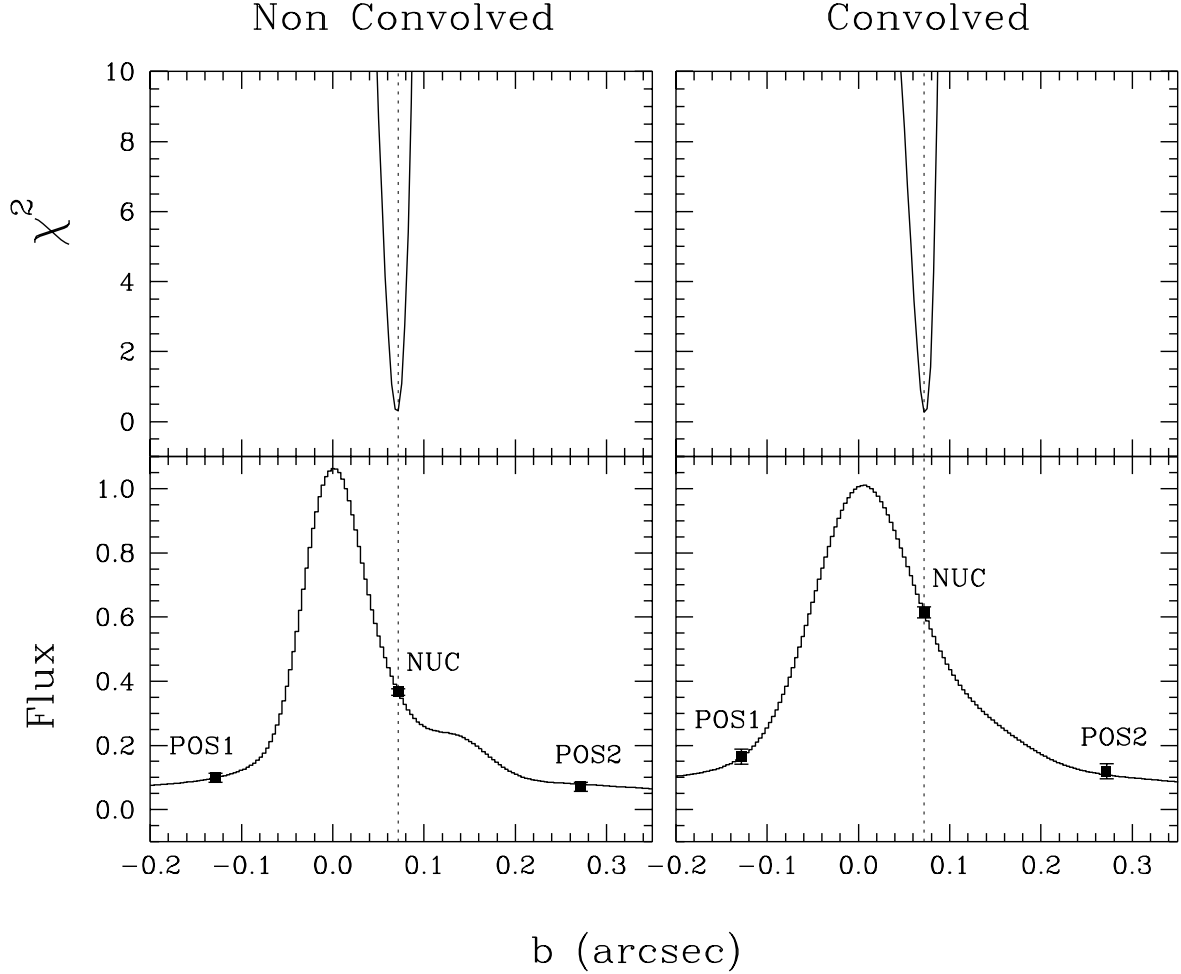,width=16cm}}
\caption{\label{fig:align} Upper panels: reduced $\chi^2$ as a function
of the impact parameter  $b$ of the central slit position (NUC). The dotted line
represents the minimum. 
Lower panels: the filled dots represent the continuum fluxes
from the spectra at NUC, POS1 and POS2 corresponding to  the impact parameter
which gives the minimum $\chi^2$; the filled lines are the
normalized luminosity profiles
in the direction perpendicular to the slit (from the
FOC f/96, F342W image).  
Right panels represent the case when the F342W image is degraded
to the f/48 spatial resolution.}
\end{figure}

A section of the F342W image, 0\farcs9 wide and parallel to the slit
orientation was extracted. Since the continuum flux measured in our
spectra is the spatial average over the slit width, we filtered the
extracted f/96 image with a flat-topped rectangular kernel 0\farcs063
wide in the direction perpendicular to the slit.  Leaving aside for the
moment the different point-spread-functions (PSF) of the f/48 and f/96, the
resulting brightness profile can be directly compared with the relative
fluxes obtained from our 
spectroscopic measurements (as shown in Fig. \ref{fig:align}).
The value of the impact parameter which best
reproduces the ratio of the fluxes measured in the three slit
positions was determined from a least--squares fit to synthetic flux
profile derived from the filter observation. As shown in Fig. \ref{fig:align}
$\chi^2$ has one well defined minimum at $b = 0\farcs07 \pm 0\farcs01$.

To take into account the possible effects of the different PSFs
for f/48 and f/96 we degraded the F342W image to the f/48
resolution.  Since the f/96 and f/48 PSFs can be
approximated by gaussians with FWHM 0\farcs04 and 0\farcs08
respectively, we convolved the f/96 image using a circularly symmetric
gaussian function with a 0\farcs07 FWHM and repeated the analysis. As before
the $\chi^2$ minimum falls at $0\farcs07 \pm 0\farcs01$
implying that the different PSFs do not
significantly affect the derived impact parameter.  The positions of the
slits with the impact parameter derived above
are displayed in Fig. \ref{fig:slitpos}, overlayed on the \ha+\nii\
continuum subtracted archival WFPC2 image .

\section{\label{sec:results} Results}

\begin{figure}
\centerline{\psfig{file=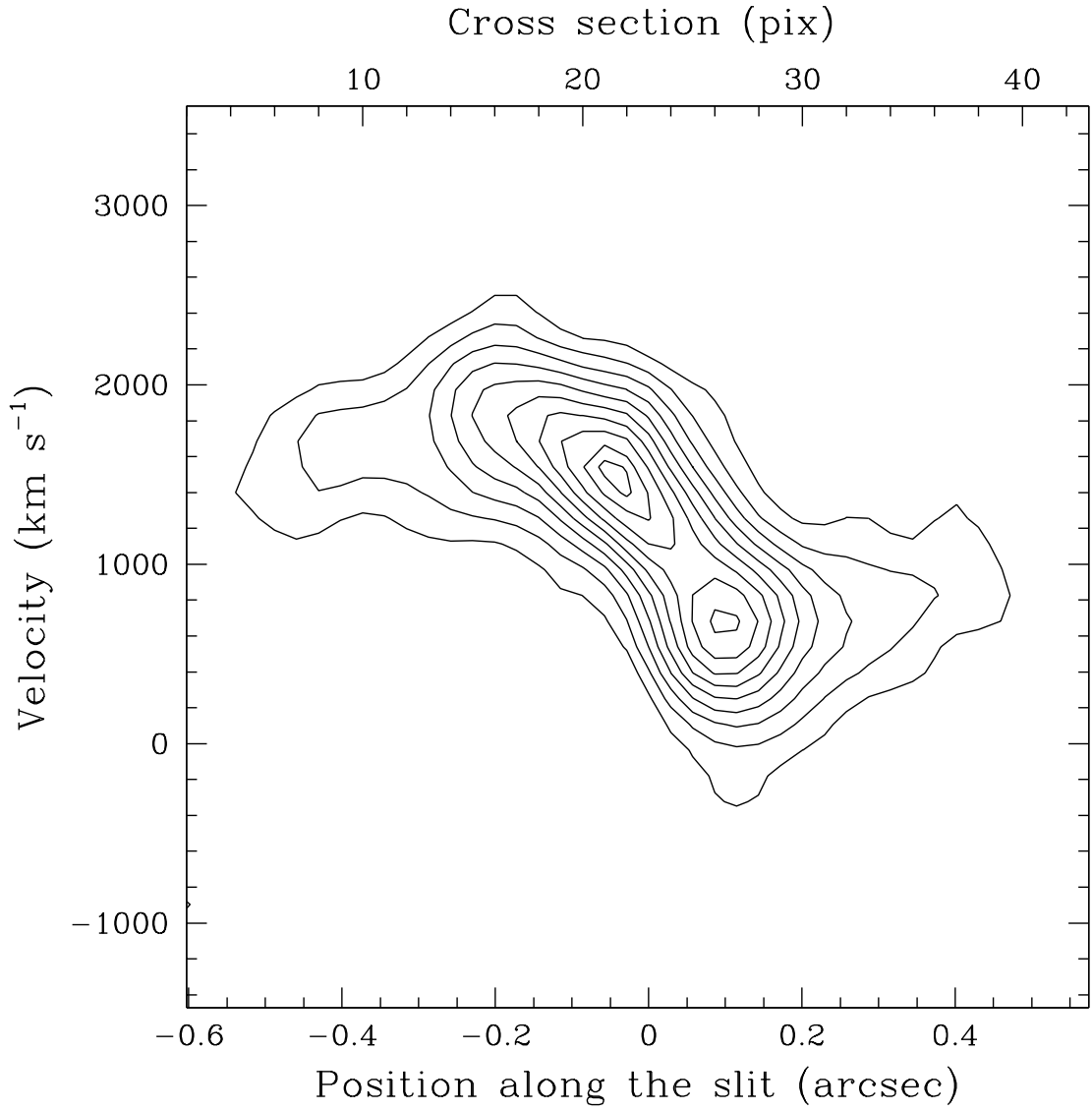,width=16cm}}
\caption{\label{fig:obscontour} Observed surface brightness contours in the 
position--velocity plane for the continuum subtracted \oii\ line at NUC. }
\end{figure}

Extended \oii\wl\wl 3726,3729 emission was detected 
in all three slit positions 
and the gray--scale, continuum subtracted \oii\ image at NUC
is displayed in Fig. \ref{fig:obscontour}.

At NUC we also detected \sii\wl\wl 4076,4069 \AA, \oiii\wl\wl 4959,5007
and possibly \hb\ emission.  Since the \oiii\ lines fall close to a
defect in the image, only the \oii\ and \sii\ were chosen
as being suitable for velocity measurements.

\begin{figure}
\centerline{\psfig{file=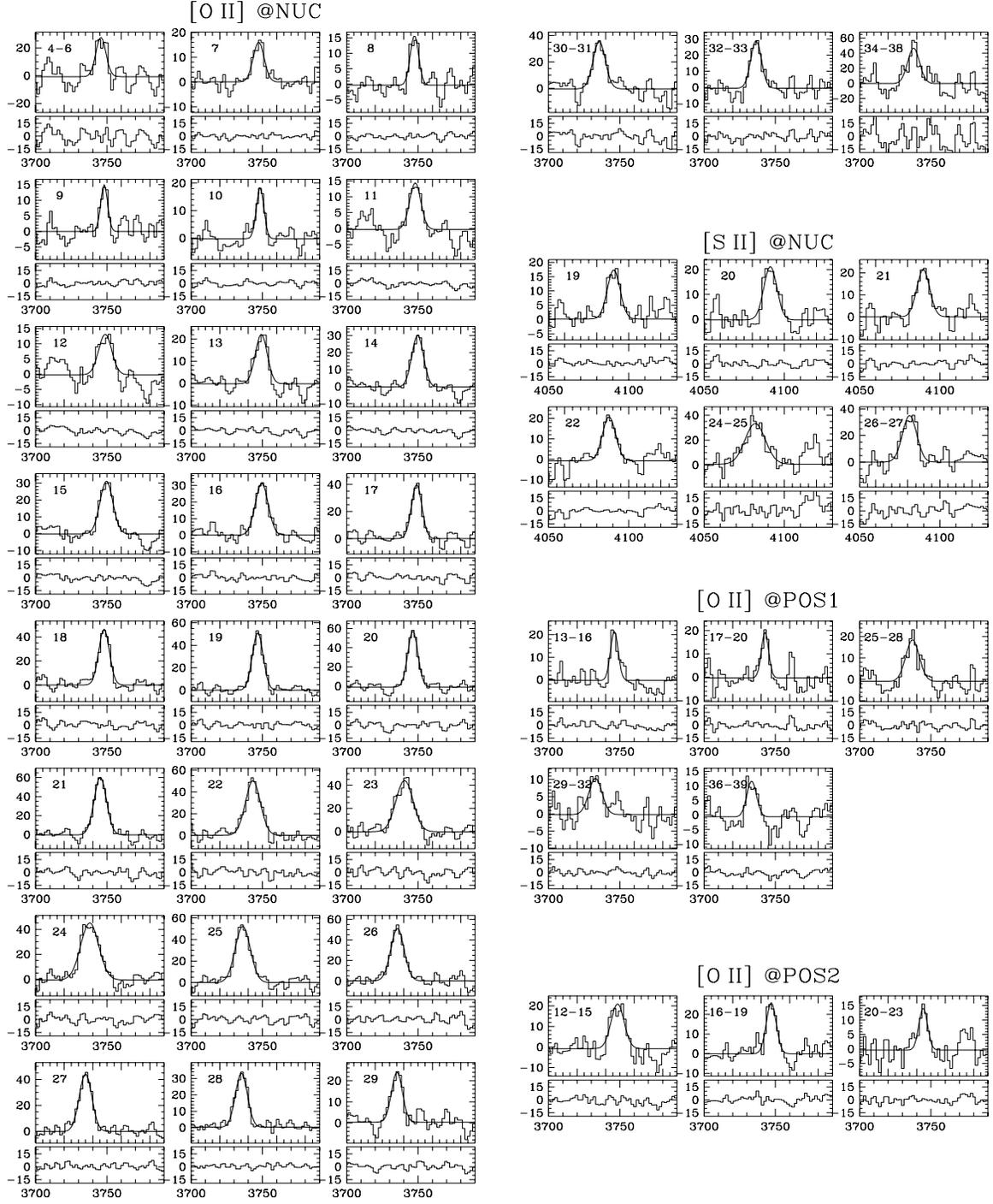,width=16cm}}
\caption{\label{fig:allfit} Observed line profiles for \oii\ (at NUC, POS1 
and POS2) and \sii\ (at NUC). The numbers in the upper left corners
of the small panels represent the single cross sections
(or the multiple ones co--added) as in the preceding
figures. The solid lines represent the single gaussian fits 
and the corresponding residuals are plotted below each panel.}
\end{figure}

The continuum subtracted lines were fitted, row by row, along the
dispersion  direction with single gaussian profiles using the task
LONGSLIT in the TWODSPEC FIGARO package (\cite{longslit}). In a
few cases, at the edges of the emission region where the signal--to--noise
ratio was insufficient, the fitting was improved by co-adding two or
more pixels along the slit direction.  All fits and respective residuals
for the \oii\ and \sii\ lines can be found in Fig. \ref{fig:allfit}.
In all cases the measured line profiles are well represented by a single
gaussian and constant continuum.

\begin{figure}
\centerline{\psfig{file=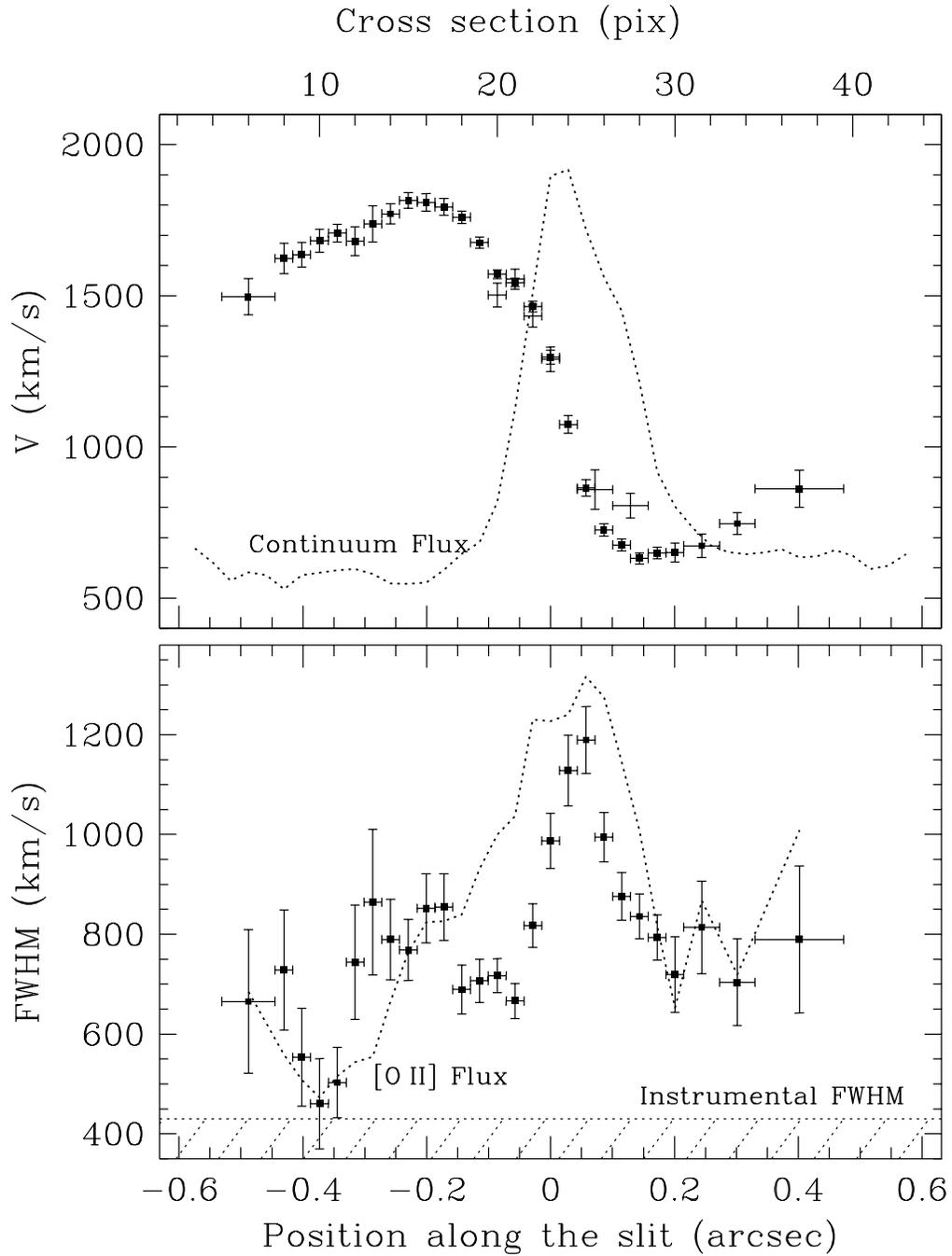,width=14cm}}
\caption{\label{fig:results1} Measured velocities and FWHMs for
\oii\ at the NUC position. The crosses in the upper panel
represent the error bars for the \sii\ measurements. The dotted
lines represent the flux distributions along the slit for the \oii\ line
and the underlying continuum.}
\end{figure}

\begin{figure}
\centerline{\psfig{file=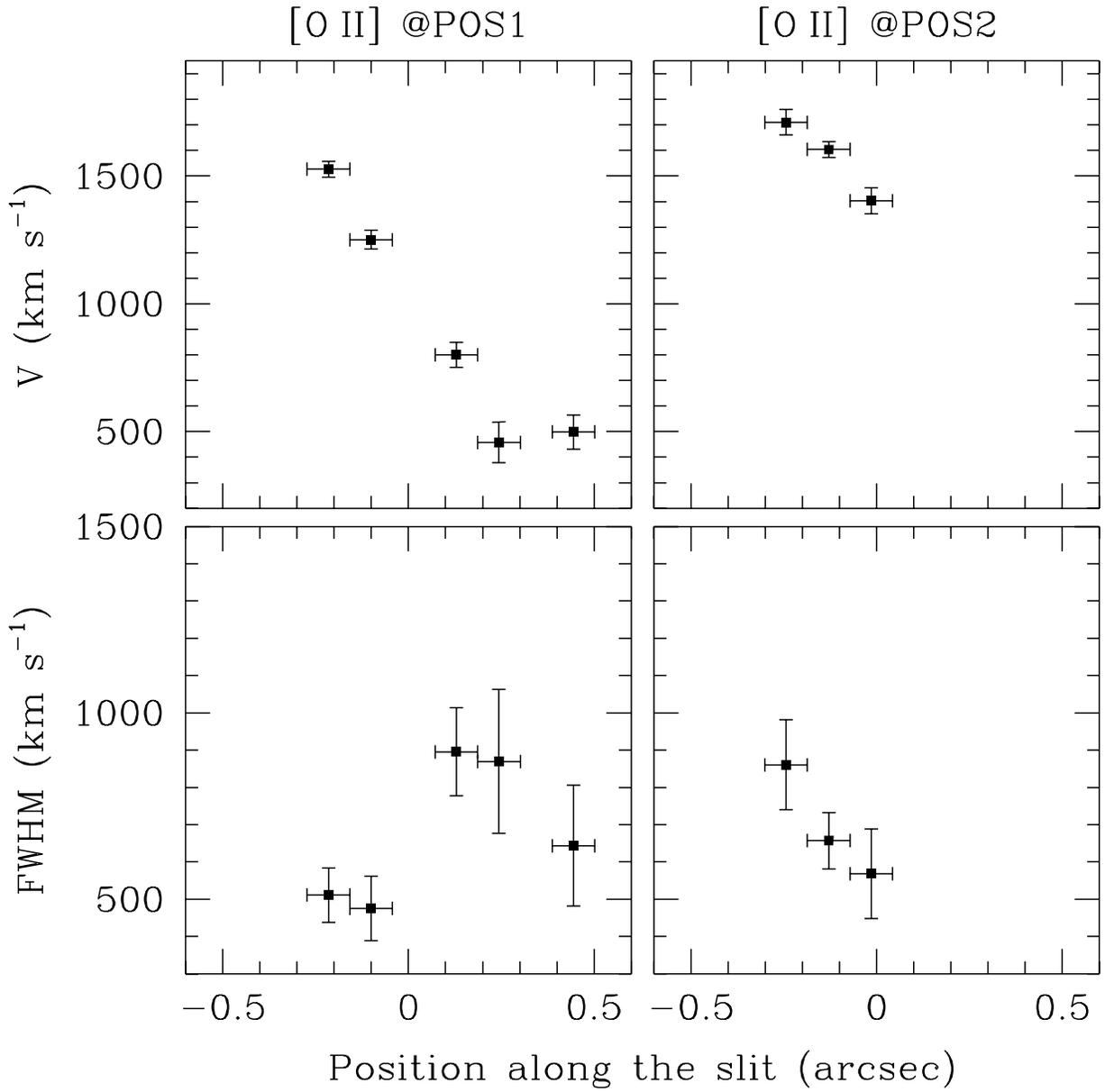,width=17cm}}
\caption{\label{fig:results2}  Velocities and FWHMs from the \oii\
line at POS1 and POS2.}
\end{figure}

The resulting central velocities, FWHM's and line intensities along the
slit are plotted in Fig. \ref{fig:results1}
for the NUC position,  and in Fig. \ref{fig:results2} for 
POS1 and POS2.
The corresponding continuum distribution along the slit is also
shown in Fig. \ref{fig:results1}.
Within the uncertainties, the velocities derived from \sii\ (thin crosses) agree
with those obtained from \oii\ (filled squares), confirming the integrity 
of the wavelength calibration.
The overall NUC velocity distribution indicates rotation with an
amplitude of 1200 km \isec, with a steep quasi--linear 
central portion between $\pm0\farcs2$ of the continuum peak and flattening
at larger radii.
Because of the much reduced signal--to--noise of POS1 and POS2
we primarily see only the brightest linear parts of the rotation curves
but we do detect a clear turn--over to the South--West of POS1 at
an amplitude of about 1000 km \isec.

The line intensity profile at NUC increases steeply toward the center
but is essentially flat-topped within the central 0\farcs14.
The width of the lines at the three slit positions is significantly
larger than the instrumental broadening (Fig. \ref{fig:allfit}),
even after taking into account  the effects of density variations on 
the wavelengths of the density sensitive doublets
(see the  discussion below).
Furthermore we note that at NUC the FWHM
and the continuum peak are shifted by $\simeq$ 0\farcs06 ($\simeq$ 2 pixels).
The significance of both these results will be discussed
in Sec. \ref{sec:profiles}.

The position--velocity diagram of the continuum subtracted \oii\ line observed
at NUC (Fig. \ref{fig:obscontour}) reveals the presence of two emission peaks
$\simeq 800$ km/sec apart in velocity, spatially separated by 
approximately 0\farcs14.
The existence of these two peaks implies that the line--emission 
does not increase monotonically to zero radius but rather
that, at a certain distance
smaller than 0\farcs07 from the nucleus ($\simeq 5$pc), the \oii\ emission
is absent.

\subsection{The impact of density variations}

\begin{figure}
\centerline{\psfig{file=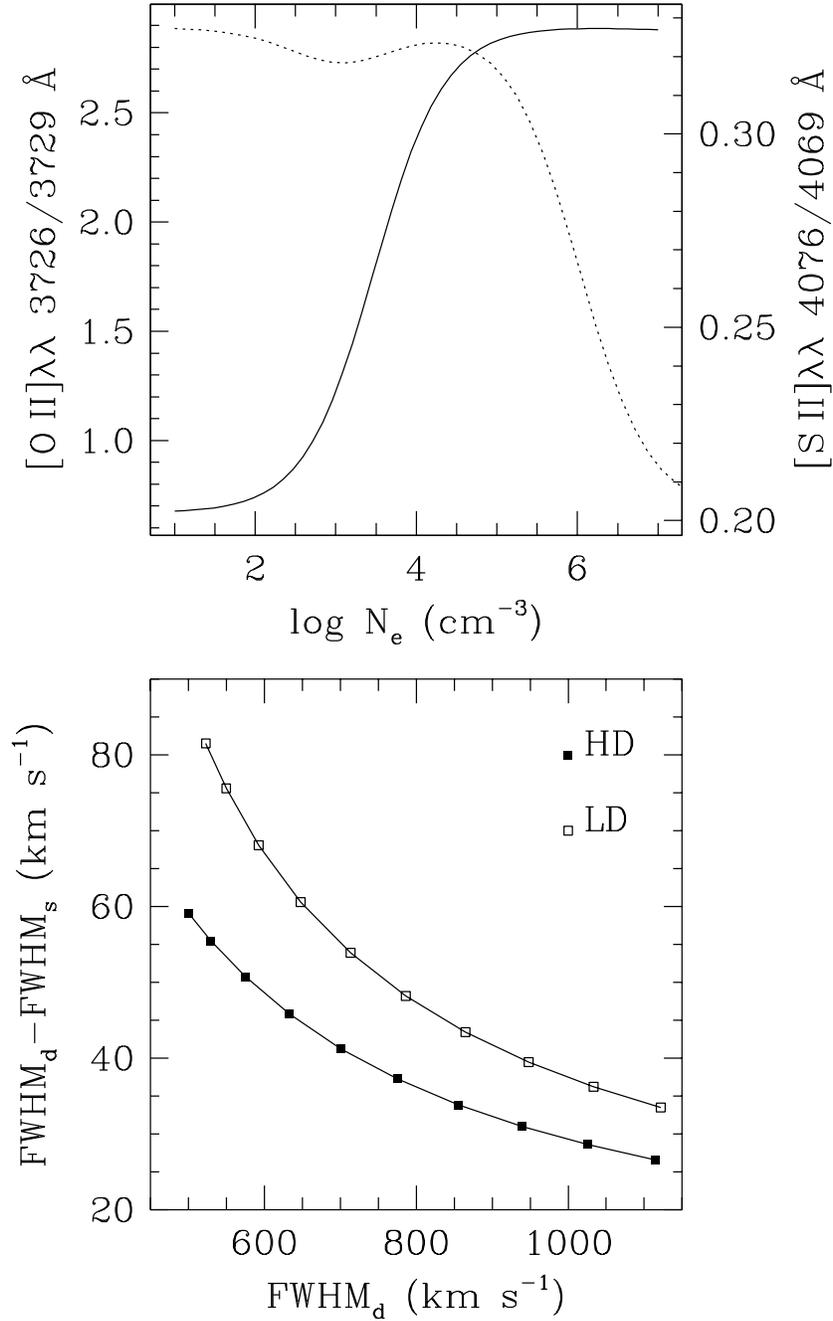,width=14cm}}
\caption{\label{fig:dens} Upper panel: Density dependence 
of the \oii\wl\wl 3726/3729 \AA\ ratio (solid line) and of the
\sii\wl\wl 4076/4069 \AA\ ratio (dotted line).
Lower panel: Broadening of the \oii\ line due to it being a doublet.
FWHM$_s$ is the width of the single line and FWHM$_d$ is the
corresponding width of the doublet. HD and LD are the high and low density
limits for the doublet, respectively.}
\end{figure}

One potential concern for the accuracy of the derived rotation curve is
that the observed \oii\ line is actually a blend of \oii$\lambda$3726.0
and \oii$\lambda$3728.8 i.e. they are separated by
225 km s$^{-1}$. We have adopted a  central wavelength of 3727.15
\AA\ in our analysis. However the doublet is density sensitive
and it is important to determine the magnitude of the shift in
the central wavelength of the doublet in response to density variations.
Using a code kindly provided by Dr. E. Oliva, we  computed the line
emissivities for the \oii\ lines using a five level atom and atomic
parameters from a compilation by \cite{mendoza}. As can be seen in Fig.
\ref{fig:dens}, the ratio between the two lines varies between 0.68 (low
density limit \Ne $<$ 50\ivol) to 2.88 (high density limit
\Ne$>$\ten{5}\ivol) and, consequently, any density--induced velocity
shift is less than $\pm$ 45 km \isec.  Furthermore, from the
archive FOS spectra described below (Sec. \ref{sec:fosdata}), the
density derived from \sii\wl 6716/\wl 6731 ranges from $\simeq$ 200 to $\simeq$
4300 \ivol\ and this restricts the range of variation to $\pm$ 25 km
\isec.  Similarly, the presence of an unresolved doublet will affect the
measurements of the line widths. The greatest effect is when the lines
are narrowest, i.e. when their FWHM is greater/equal to the instrumental
broadening ($\simeq$ 430 km \isec). In that case the FWHM of the \oii\
doublet can be $\simeq$100 km \isec\ broader than that of a single
line.  When the FWHM is larger than 600 km \isec\ the broadening is less
than 60 km \isec\ i.e. negligible for our data (Fig. \ref{fig:dens}).

We applied a similar treatment to the density sensitive \sii\
doublet (Fig. \ref{fig:dens}) deriving a central wavelength of 4070.2 \AA\
with a range of variation of $\pm$ 25 km \isec\
(atomic data from \cite{pradhan}).
If, as the above density measurements imply, we are in the low density
limit for the \wl 4076/\wl 4069 doublet then the central
wavelength would be shifted to 4070.5 \AA, implying an uncertainty
of at most 25 km \isec\ in our assumed rest wavelength.
We conclude that the variations induced by density effects are always within
our measurement uncertainties.

\section{\label{sec:fosdata} Comparison with archival FOS observations}

\begin{figure}
\centerline{\psfig{file=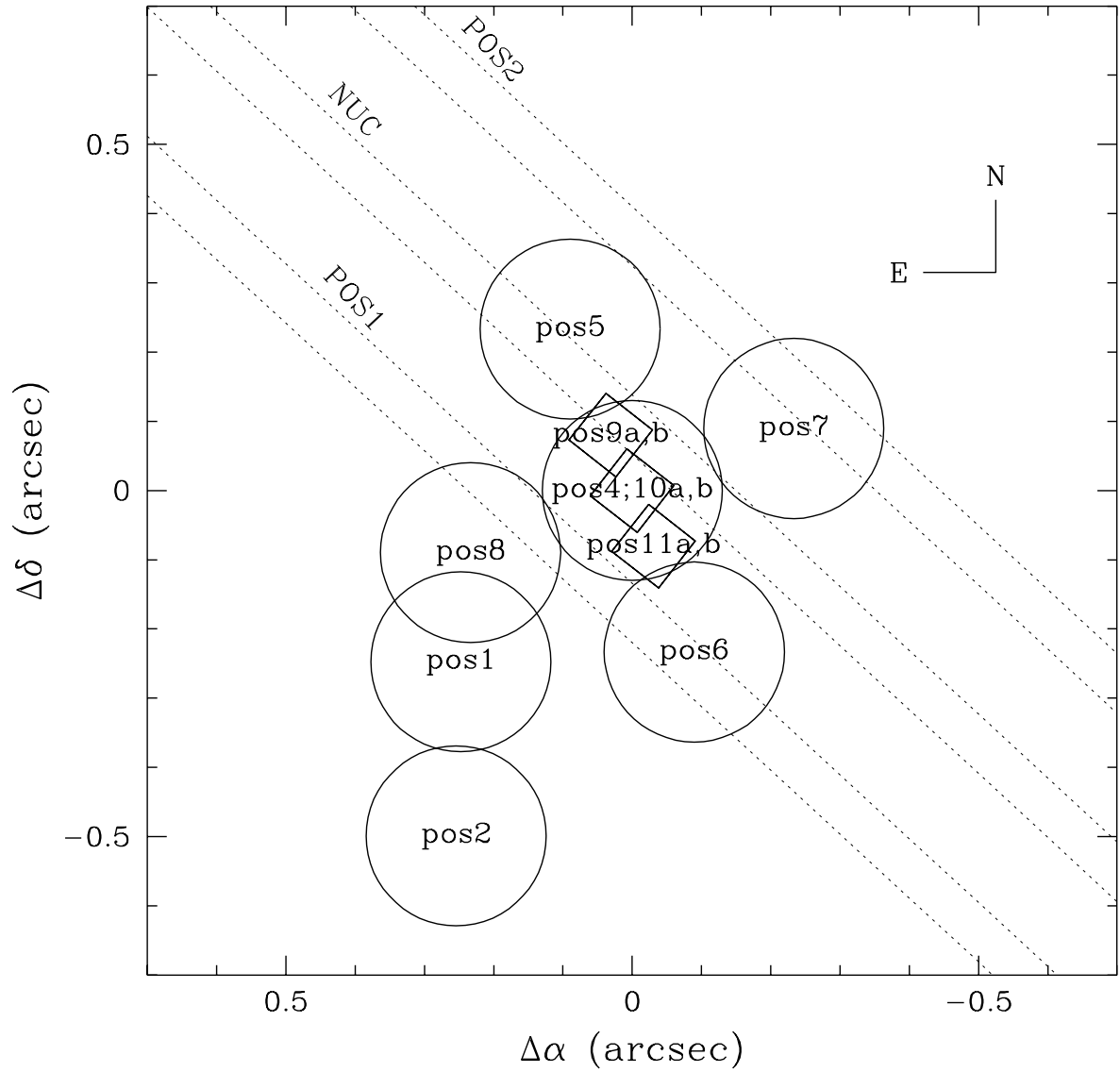,width=16cm}}
\caption{\label{fig:focfos} Location of FOS apertures and
corresponding sizes compared with
the slit positions and the FOC long--slit observations.}
\end{figure}

We retrieved the FOS data used by \cite{fos94} and \cite{fos96}
from the HST archive. The datasets used and the corresponding
target names are listed in Table 2, according to their  
notation, and their positions are compared with those of our slits in
Fig.~\ref{fig:focfos}.

\begin{figure}
\centerline{\psfig{file=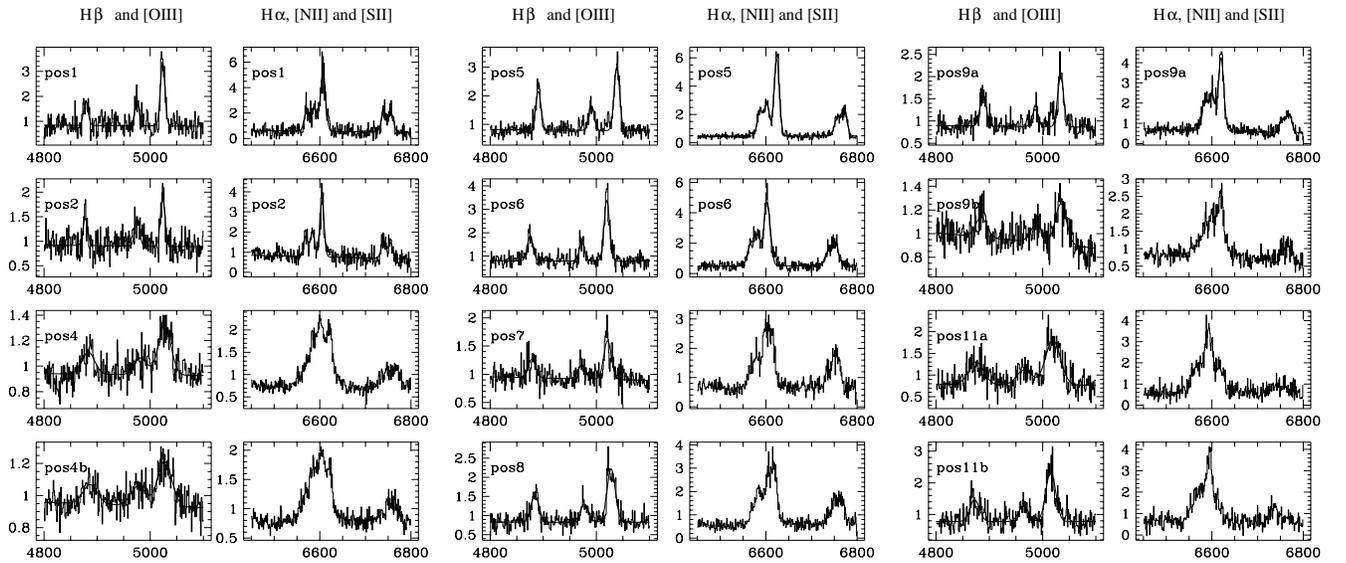,height=8cm,angle=-90}}
\caption{\label{fig:fosres} Original, non smoothed FOS data around
\oiii\ and \ha+\nii.
The position labels are those of the preceding figure. The bold, solid line
represent a one component gaussian fit (when possible).}
\end{figure}

Because the FOS data are rather noisy we smoothed them by convolution
with a 1.8\AA--sigma gaussian,
i.e. the FOS spectral resolution, and then 
measured the velocities of \hb, \oiii\
\oi, \ha, \nii\ and \sii\ using single--gaussian fits
as shown in Fig. \ref{fig:fosres}. 
When possible \hb\ and the two \oiii\ lines were fitted simultaneously
under the constraint that they had the same FWHM.
In some cases it was also possible to satisfactorily deblend \ha,
\nii\ and the \sii\ doublet.
The measured ratio of the \sii\ doublet implies
an electron density which varies between 200 \ivol\ and 4300 \ivol. 
The measured velocities are given in Table 3
and are in acceptable agreement with the values given in Table 1 of
\cite{fos94}. The similarity between the velocities obtained from 
different ionic
species indicates that our results are not unduly biased by variations in 
the ionization conditions in the disk.

\begin{figure}
\centerline{\psfig{file=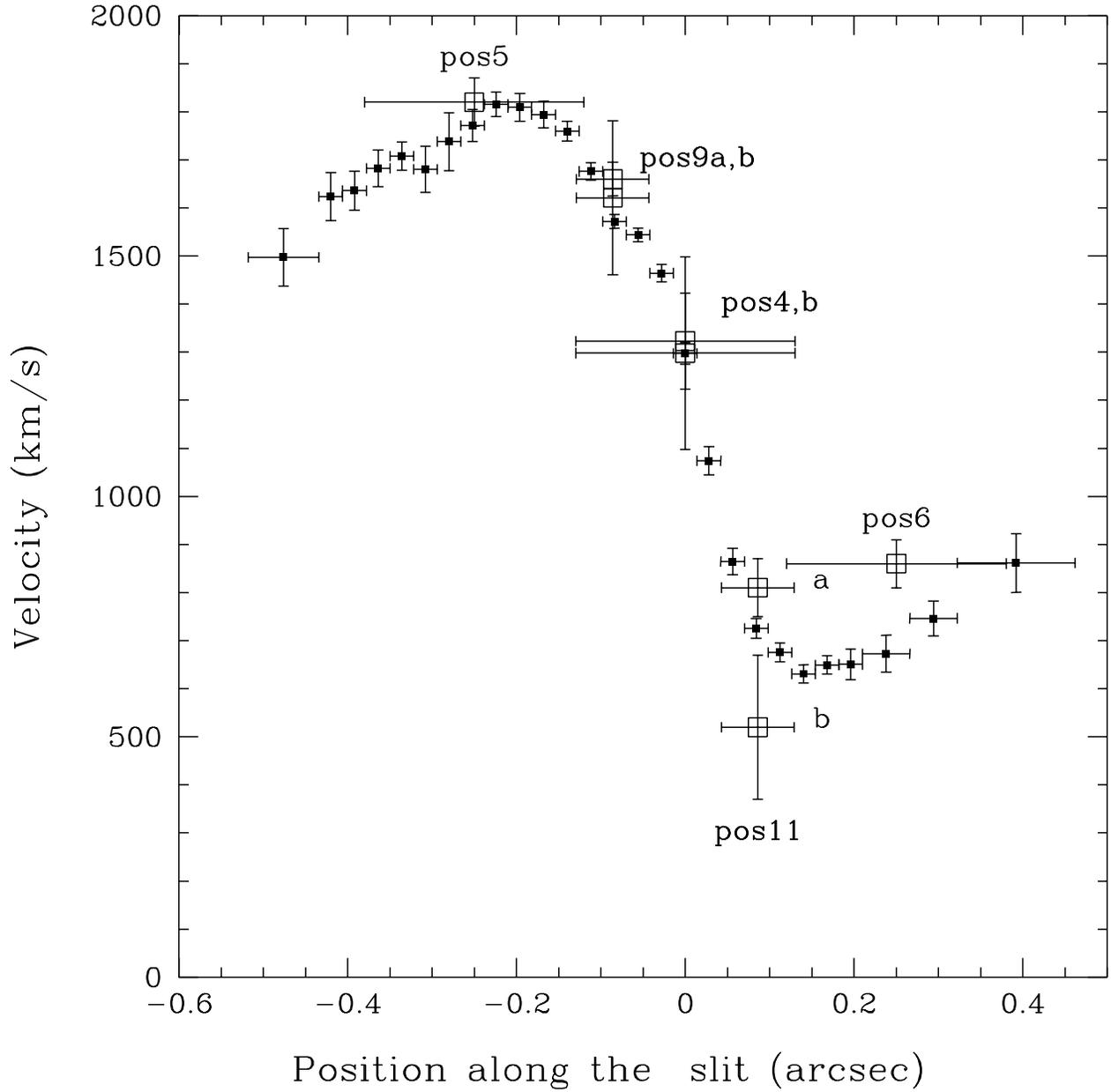,width=17cm}}
\caption{\label{fig:comp} The archival FOS data (empty squares) are
compared with the \oii\ FOC rotation curve (filled squares). }
\end{figure}

In Fig. \ref{fig:comp} we compare the velocities obtained from
our slit position NUC
with those obtained from the FOS at POS4 through 6, in the
0\farcs26 aperture and at POS9 through 11, in the smaller 0\farcs09
aperture. The plotted uncertainties of the FOS data,
typically between 50 and 100 km \isec\ in a given
dataset,  are  the internal scatter of  the
velocities measured in a given aperture.
Within the substantially larger uncertainties, the FOS rotation curve is in
reasonable accord with our results.
It is also clear that our data represent a considerable
improvement in both spatial resolution and accuracy in 
the determination of the rotation curve of the disk.

\section{\label{sec:fit} Modeling the rotation curve: a simple
but constructive approximation}

We now derive the expected velocity measured along the slit for
a thin disk in keplerian motion in a gravitational potential dominated by
a condensed central source. 
At this stage
we ignore both the finite width of the slit and effects of the
PSF which will be considered in the next section.
Although the limitations of this approximation are clear, since the angular 
scale of the region of interest is similar to the size of 
point--spread--function, 
several general conclusions can be drawn from this simplified
treatment which clarify the behaviour or the more realistic model
fits described in Sec. 8.
For simplicity we will also refer to the condensed central
source as a black hole deferring the reality of this
assumption to Sec. 10.

Any given point $P$, located on the disk at a radius $R$, 
has a keplerian velocity 
\begin{equation}
V(R) = \left(\frac{GM_{BH}}{R}\right)^\frac{1}{2}
\end{equation}
where $M_{BH}$ is the mass of the black hole.

\begin{figure}
\centerline{\psfig{file=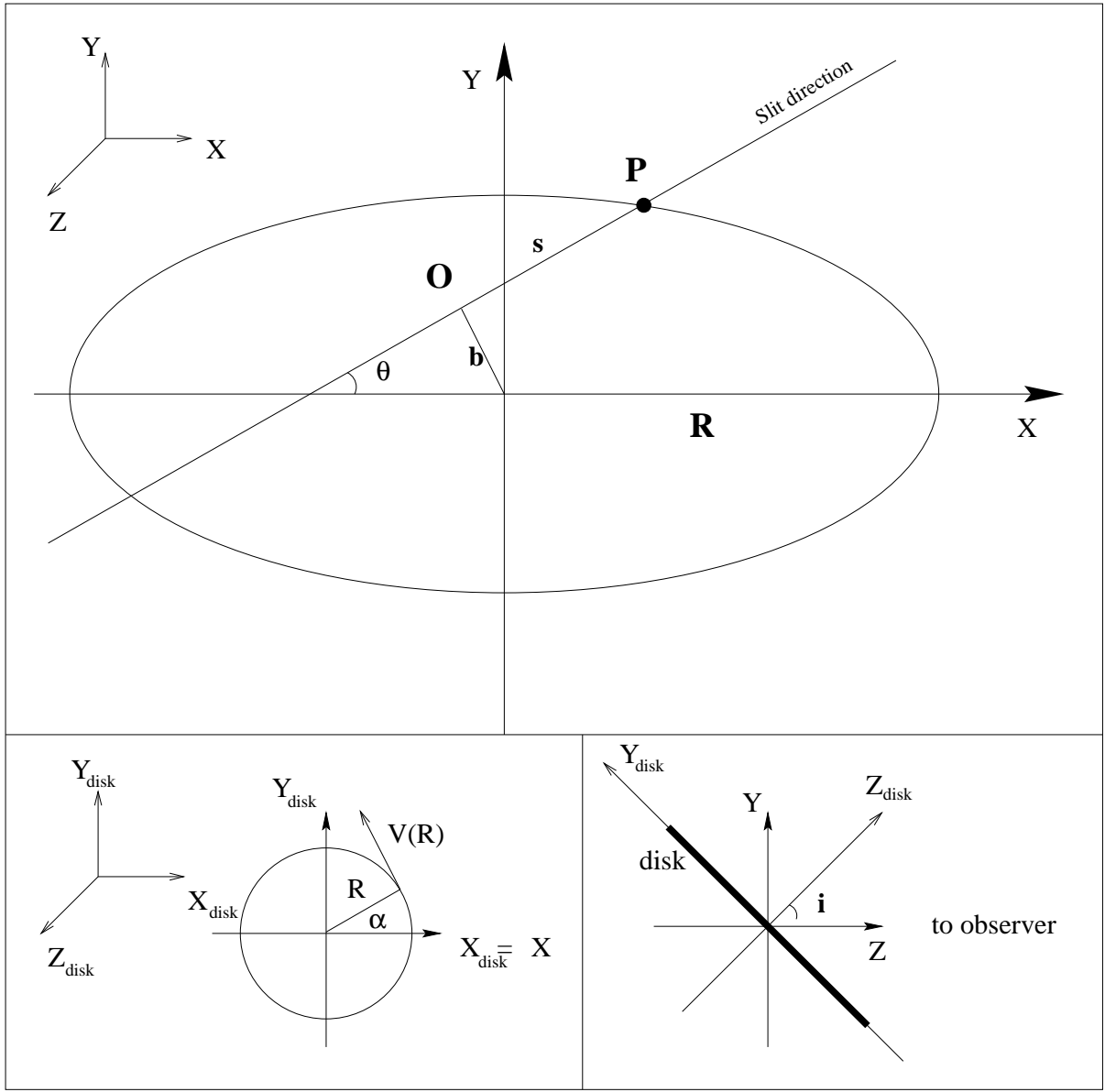,width=16cm}}
\caption{\label{fig:disk} Schematic representation of the reference
frames used in the determination of the keplerian rotation curve.
$XY$ is the plane of the sky, $X$ is directed along the major axis of the disk;
$Z$ is directed toward the observer. $X_{disk}Y_{disk}$ is
the reference frame on the disk plane such that $X_{disk}=X$.}
\end{figure}

We choose a reference frame such that the X and Y axis,
as seen on the plane of the sky, are along
the major and minor axis of the disk respectively (see Fig. \ref{fig:disk}).
In this coordinate system a point $P(X,Y)$ is at a radius $R$
such that
\begin{equation}
\label{eq:eq1}
X^2+\frac{Y^2}{(\cos i)^2 } = R^2
\end{equation}

Each point along the slit can
be identified by its distance {\bf s} to the ``center'' of the slit {\bf O}, 
whose distance from the nucleus defines the ``impact parameter'' $b$.
Let $\theta$ be the
angle between the slit and the major axis of the disk, i.e. the line of nodes,
and define $i$ to be the inclination of the disk with respect to
the line of sight.
Since $P(X,Y)$ is located on the slit, X and Y are given by
\begin{equation}
\label{eq:eq2}
X = -b\sin\theta+s\cos\theta \\
\end{equation}
\begin{equation}
\label{eq:eq3}
Y = b\cos\theta+s\sin\theta \\
\end{equation}

The circular velocity $V(R)$ is directed tangentially to the disk as
in Fig. \ref{fig:disk} and its projection along the line of sight is then
$-V(R)\cos\alpha \sin i$ (the - sign is to take into account the
convention according to which blue--shifts result in negative velocities).

Making the transformation between coordinates on the plane of the disk
and the X,Y on the plane of the sky
\begin{equation}
\tan\alpha = \frac{Y}{X\cos i}
\end{equation}
hence, if $V_{sys}$ is the systemic velocity,
the observed velocity along the slit is given by
\begin{equation}
\label{eq:kepdisk}
V = V_{sys}-(GM_{BH})^\frac{1}{2}
\frac{(\sin i)\, X}
{\left( X^2+\frac{Y^2}{(\cos i)^2}\right)^\frac{3}{4}}
\end{equation}

A non--linear least squares fit of 
the model defined in the equation \ref{eq:kepdisk}, with
$M_{BH} (\sin i)^2$, $\theta$, $i$, $b$, $V_{sys}$ and
$cpix$ (which defines the origin of the $s$ axis) as
free parameters to the observed rotation curve was 
obtained using simplex optimization  code.
Since the error bars on the independent
variable are not negligible, we took them into account by minimizing
the modified norm:
\begin{equation}
\label{eq:chisq}
\chi^2 = \sum_{i=1}^N \frac{\left(y_i-V(x_i; p_1, ..., p_6)\right)^2}
  {\Delta y_i^2+\left(\frac{\partial V}{\partial x}\right)_{x_i}^2
  \Delta x_i^2}
\end{equation}
where $y_i$ is the measured velocity at the pixel $x_i$, there
are $N$ data points and $p_i$ are the free parameters of the fit.

Because of the complexity of the fitting function
we also carried out many trial minimizations using different
initial estimates for the most critical free parameters,
i.e. $i$, $\theta$ and $b$, taken from a large grid 
of several hundred, evenly spaced values.
Many local minima of the $\chi^2$ function were found and we only accepted 
those solutions with a reduced $\chi^2<2.5$ and an
impact parameter, $b$, consistent with that determined
in Sec. \ref{sec:impact} ($0\farcs06<b<0\farcs08$).

Even though there is considerable
non--axisymmetric structure in the data,
in their original analysis \cite{fos94} used a value for the
inclination, $i=42^\circ\pm5$, determined from isophotal fits
to the surface photometry of the disk at radii
ranging between 0\farcs3 and 0\farcs8
from the nucleus.
Unfortunately, from our analysis of the imaging data it is not clear
whether this result, obtained on the large scale, is valid at small
radii (the inner 0\farcs3). 
To determine the inclination on the basis of surface photometry of
the emission lines, 
higher spatial resolution images are needed and this
can only be accomplished by HST measurements at UV
wavelengths.
Even then the bright non--thermal nucleus may dominate the structure
of the central region.
Indeed from our analysis of the kinematics the inclination is the most poorly
constrained parameter, with
acceptable values ranging from 47$^\circ$ to 65$^\circ$.
Though the other parameters are intrinsically rather
well constrained,
in Table 4 we therefore show the variation of  their allowed values 
for two inclination ranges.

A few points are evident from this analysis:
the small angle $\theta$ to the line of nodes
($-5^\circ<\theta<4^\circ$) is a consequence
of the apparent symmetry of the two branches of the rotation curve.
When the impact parameter is non--null this symmetry can be achieved only
if the slit direction is close to that of the line of nodes.
The center of the rotation curve (between pixels 22.6 and 22.9)
is close to the peak of the continuum distribution along the slit
(pixels $\simeq$23--24) as one would expect if the latter indicates the
location closest to the nucleus. However one would also expect 
the point with largest FWHM (pixel 25) to be coincident with it
and not to be shifted by $\simeq$ 2 pixels ($\simeq0\farcs06$),
as observed.
We will return to this issue in the following sections.
The systemic velocity is in reasonable agreement
with the estimate of 1277 km \isec\ determined by van der Marel (1994)
when one takes into account the uncertainty on the zero point of the wavelength 
calibration.

\begin{figure}
\centerline{\psfig{file=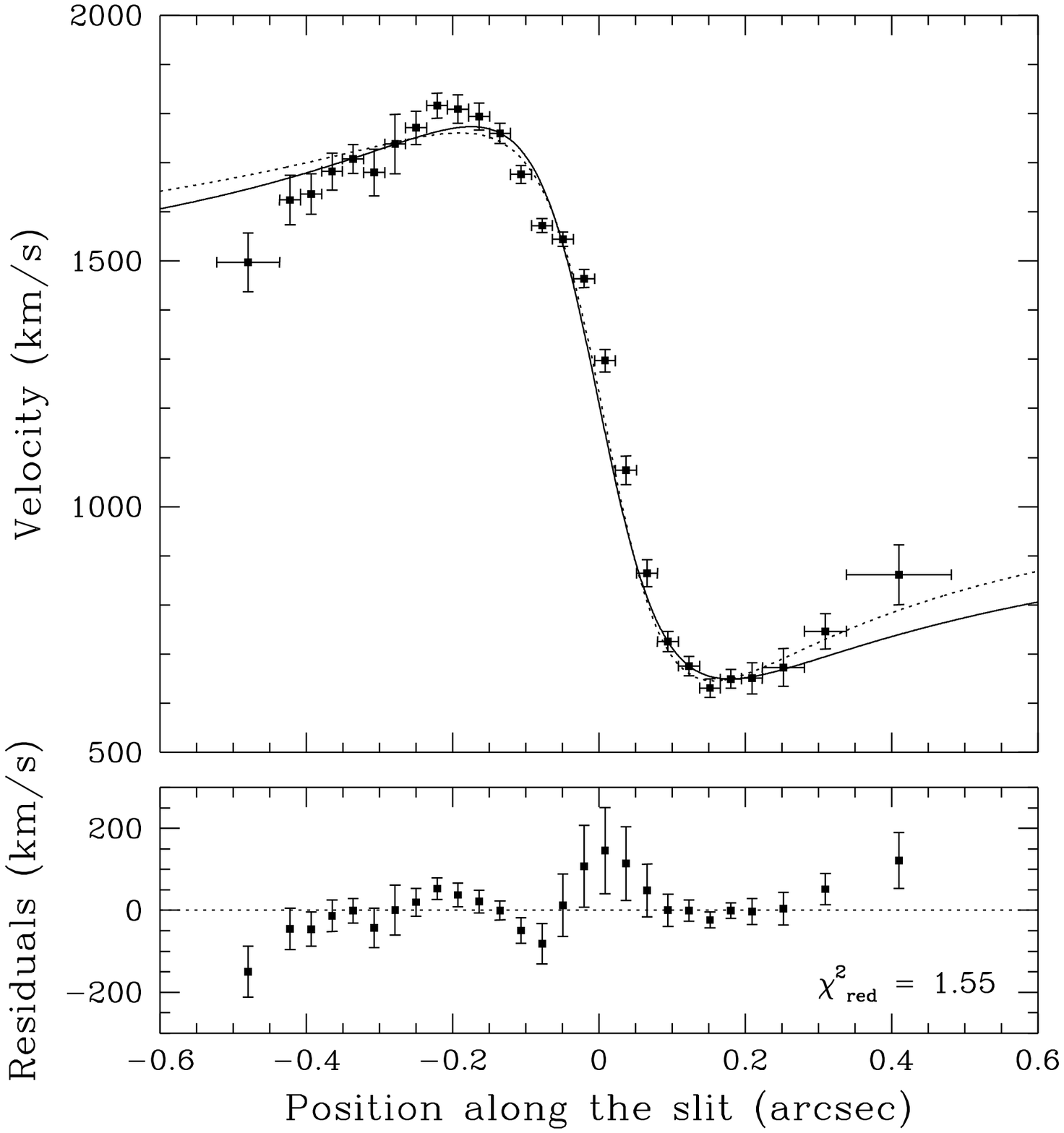,width=17cm}}
\caption{\label{fig:bestfit1} Best fits of the observed rotation curve 
with the keplerian thin--disk model.
The solid line corresponds to 
$cpix=22.7$, $b$=0\farcs08, $M_{BH} (\sin i)^2$=1.73\xten{9}\Mo,
$\theta=0.7^\circ$, $i=49^\circ$ $ V_{sys}$=1204 km \isec\ and
the dotted line to $cpix=22.7$, $b$=0\farcs06,
$M_{BH} (\sin i)^2$=1.68\xten{9}\Mo,$\theta=-5.1^\circ$,
$i=60^\circ$ $ V_{sys}$=1274 km \isec\ ($\chi^2=1.73$).
The residuals are computed
for the former set of values and the error-bars on the velocity are the
square roots of the denominators of equation 9.}
\end{figure}

The two fits which have the minimum and maximum of the acceptable 
$\chi^2$ values  are shown in Fig. \ref{fig:bestfit1} as solid and
dotted lines, respectively, and the corresponding values of the
parameters are:
$cpix=22.7$, $b$=0\farcs08, $M_{BH} (\sin i)^2$=1.73\xten{9}\Mo,
$\theta=0.7^\circ$, $i=49^\circ$ $ V_{sys}$=1204 km \isec\
($\chi^2$=1.55) and
$cpix=22.7$, $b$=0\farcs06, $M_{BH} (\sin i)^2$=1.68\xten{9}\Mo,
$\theta=-5.1^\circ$, $i=60^\circ$ $ V_{sys}$=1274 km \isec
($\chi^2$=1.73).
Note that the error bars in the plot of the residuals
are the square roots of the denominators in equation \ref{eq:chisq}.

Taking into account all possible fits which are compatible with the
data, the preliminary estimated value for the projected 
mass is $M_{BH} (\sin i)^2=1.7^{+0.2}_{-0.1}\xten{9}$\Mo
and $M_{BH} =2.7\pm0.5\xten{9}$\Mo\ with the allowed range
of variation in $i$. In later sections we will derive a more
accurate value for this important parameter.

\section{\label{sec:fitpsf}The smearing effects of the PSF}

While the results of the preceding section give a reasonably 
satisfactory fit to the observed rotation
curve by assuming a simple keplerian model, there are two significant
issues which need to be addressed. Inspection of
Fig. \ref{fig:results1} reveals that: i) the lines are broad in
the inner region (FWHM$>1000$ km \isec) and ii) the broadest lines do
 not occur at the center of rotation but at a distance of $\simeq 0\farcs06$ (2 pixels)
from it.
Taken in conjunction these facts may imply that the gas disk is not in 
keplerian rotation which could potentially invalidate any derived mass estimate.
So far we have ignored the combined
effects of the PSF, the finite slit--width and the intrinsic
luminosity distribution of the gas. In this section we include these
effects in our analysis and this enables us to reconcile these worrying features
of the gas kinematics with the keplerian disk.

To take into account the effects of the f/48 PSF
and the finite slit size we must average the velocities using the luminosity
distribution and the PSF as weights. To compute the model curve we 
chose the reference frame described by $s$ and $b$ i.e.
the coordinate along the 
slit and the impact parameter. With this choice 
the model rotation curve $V_{ps}$ is given by the formula:
\begin{equation}
V_{ps}(S) =
\frac{\int_{S-\Delta S}^{S+\Delta S} ds \int_{B-h}^{B+h} db
\int\int_{-\infty}^{+\infty} db^\prime ds^\prime
V(s^\prime,b^\prime) I(s^\prime,b^\prime) P(s^\prime-s, b^\prime-b) }
{ \int_{S-\Delta S}^{S+\Delta S} ds \int_{B-h}^{B+h} db
\int\int_{-\infty}^{+\infty} db^\prime ds^\prime
		    I(s^\prime,b^\prime) P(s^\prime-s, b^\prime-b) }
\end{equation}
where $V(s^\prime,b^\prime)$ is the keplerian velocity derived in eq. \ref{eq:kepdisk},
$I(s^\prime,b^\prime)$ is the intrinsic luminosity distribution of the
line, $P(s^\prime-s, b^\prime-b)$ is the spatial PSF of the f/48
relay along the slit direction.
$B$ is the impact parameter (measured at the center of the slit)
and $2h$ is the slit size, S is the position along the slit at which
the velocity is computed and $2\Delta S$ is the pixel
size of the f/48 relay.
For the PSF we have assumed a gaussian with 0\farcs08 FWHM i.e.
\begin{equation}
P(s^\prime-s, b^\prime-b) = \frac{1}{\sqrt{2\pi\sigma^2}} \exp \left(
-\frac{1}{2}\frac{(s^\prime-s)^2}{\sigma^2}
-\frac{1}{2}\frac{(b^\prime-b)^2}{\sigma^2} \right)
\end{equation}

\begin{figure}
\centerline{\psfig{file=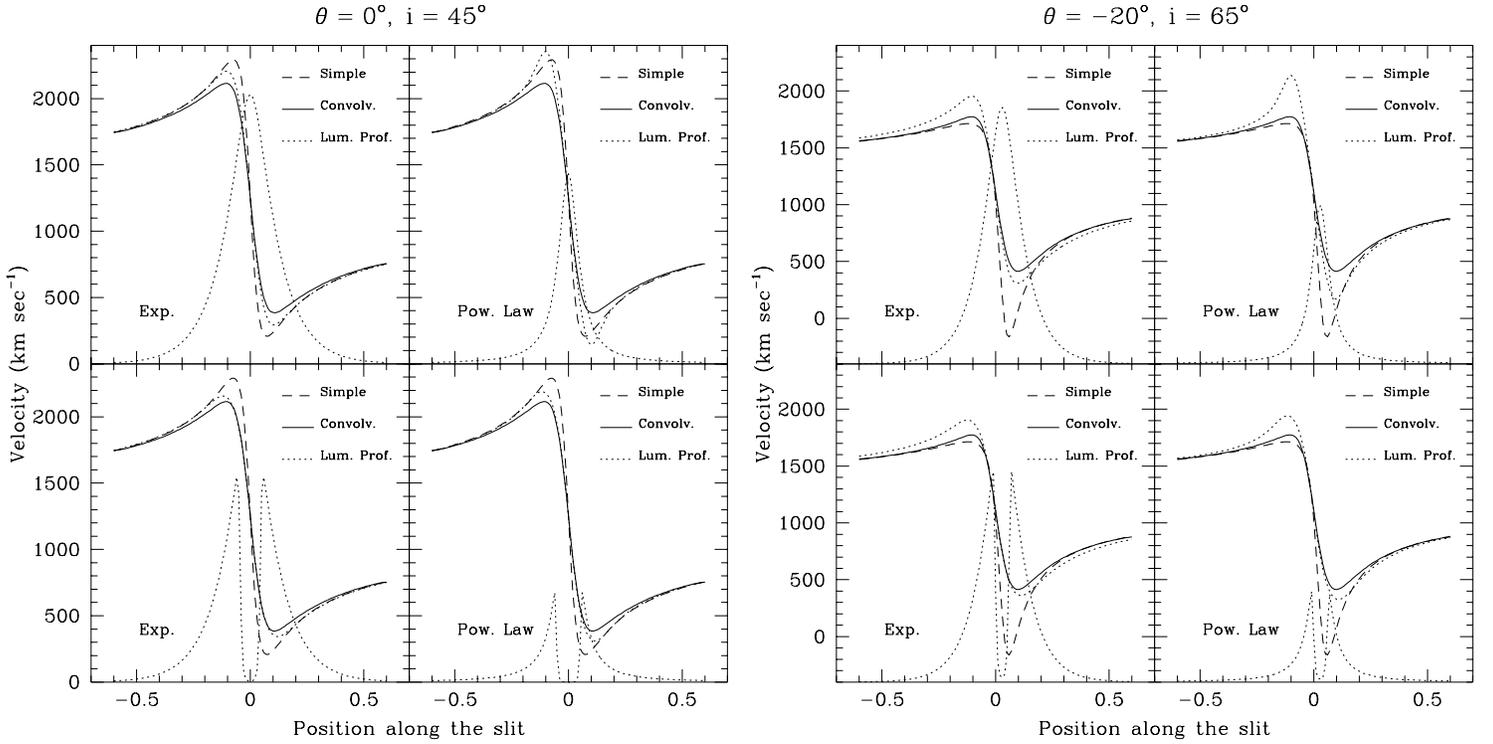,height=10cm,angle=-90}}
\caption{\label{fig:rotpsf} Model rotation curves computed from the 
equations of Sec. 7 and 8 and representing ``extreme'' cases i.e.
with parameters chosen to maximize the various effects.
For all the curves $b=0\farcs03$ and $M_{BH}(\sin i)^2=2.5\xten{9}\Mo$,
$\theta$ and $i$ are indicated at the top of each panel.
The dashed line is the rotation curve from the analytical formula of
Sec. 7. The solid line is derived from the simple
convolution of the analytical relation with the 2D spatial PSF.
The dotted line takes into account the luminosity distribution which
is also plotted with the same line style. ``Exp.'' and ``Pow. Law''
are the exponential and power law luminosity distributions 
described in Sec. 9. In the lower inserts in both panels these
luminosity distributions are multiplied by a ``hole'' function (Sec. 9).
The effects of averaging over the slit--width are always
so small as to not being visible in this figure.}
\end{figure}

The consequences of this more realistic approach to modeling the rotation curve
are illustrated in Fig. \ref{fig:rotpsf} for some extreme choices of
both the luminosity distribution
(power law or exponential profile) and the geometric parameters
of the disk.
Motivated by the presence of two peaks in the position velocity of figure \ref{fig:obscontour}
we also included a case in which the line emission is absent in the very center of the disk.
In each case we show the importance of the convolution
with the spatial PSF and the weighted average with luminosity profile
and slit width.
In general the dominant effect on the 2D velocity field is the convolution 
with the spatial PSF, and since the slit width is narrower than the PSF
it has little or no effect in modifying the expected rotation curve.
The effects of the luminosity distribution are important only when the curve
is strongly asymmetric with respect to the center of rotation
i.e. when the impact parameter is not null and the angle with the line
of nodes is much greater than zero. These effects are larger for steeper
luminosity distributions and lead to large
velocity excursions from the PSF--convolved velocity field at the turn--over
radii (see Fig. \ref{fig:rotpsf}, right panel).
Fortunately, these extreme cases can be eliminated from further discussion 
because they are not a good representation of the observed rotation curve
for M87. In the cases of interest,
the differences at the turn--over radii are always less
than $\simeq$100 km \isec\ and neglecting the 
weighting of the luminosity distribution
can result in an over--estimate of the mass of up to 0.5\xten{9}\Mo,
still within the formal uncertainties of the fit derived below.
The weak dependence of the model rotation curve
on the luminosity distribution is important because the true luminosity
distribution for \oii\ is unknown.

The presence of a ``hole'' at the center of the luminosity
distribution whose size is comparable with the FWHM of the PSF
has little effect on the rotation curve but, as we shall see in
Sec. 9, holes do have an effect on the width of the line profiles.

Using this modified fitting function under the same basic assumptions
described in Sec. \ref{sec:fit} leads to the parameters given in Table 4.
The errors quoted are conservative as they are based on the mean absolute 
deviation  of values obtained from the histogram of the local minima.

As a sanity check, we have repeated the above fitting procedure
taking into account the luminosity profiles plotted 
in Fig. \ref{fig:rotpsf} i.e. exponential and power
law dependences on radius (see Sec. \ref{sec:profiles}). 
We found that the luminosity weighting introduces no significant change
in the loci of acceptable solutions.

The PSF smearing has three effects on the model fits, firstly,
as one would expect, the required black hole mass is increased 
to compensate for the lowering of the velocity amplitudes.
In addition the inclination is more poorly constrained and larger
angles with the line of nodes become admissible.
However, taking into account the POS1 and
POS2 data restricts the inclination to less than $\simeq65^\circ$.

\begin{figure}
\centerline{\psfig{file=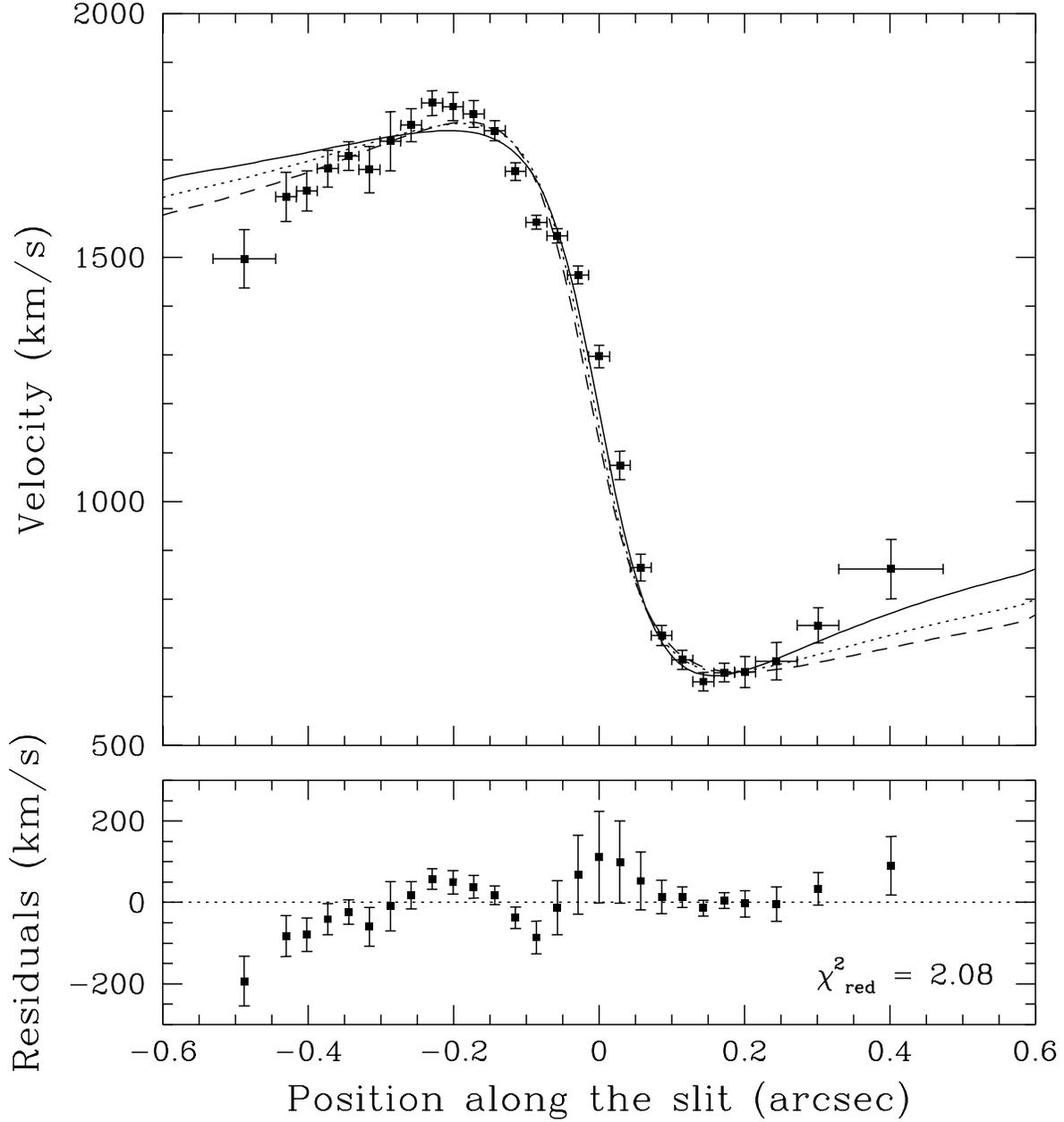,width=17cm}}
\caption{\label{fig:bestfit2} Best fits to the observed rotation curve
taking into account the smearing due to the spatial PSF; the solid
line correspond to the parameters of fit A, the dotted line to fit B and the
dashed line to fit C (see sec. 8).
The errors on the residuals are as in Fig. 12.}
\end{figure}

Three representative fits with acceptable values of the reduced $\chi^2$ 
are shown in  Fig. \ref{fig:bestfit2} and have the corresponding parameters:
fit A: $cpix=23.0$, $b$=0\farcs08, $M_{BH} (\sin i)^2$=1.91\xten{9}\Mo,
$\theta=-9^\circ$, $i=51^\circ$ $ V_{sys}$=1290 km \isec\
($\chi^2$=2.08), 
fit B: $cpix=22.7$, $b$=0\farcs08, $M_{BH} (\sin i)^2$=1.93\xten{9}\Mo,
$\theta=1^\circ$, $i=52^\circ$ $ V_{sys}$=1203 km \isec
($\chi^2$=1.90) and
fit C: $cpix=22.5$, $b$=0\farcs085, $M_{BH} (\sin i)^2$=2.00\xten{9}\Mo,
$\theta=7^\circ$, $i=50^\circ$ $ V_{sys}$=1146 km \isec\
($\chi^2$=1.82). 
The main difference between the fits is in the sign of 
$\theta$ since the analysis of the rotation curve alone cannot
distinguish between them but, as described in the next section (9),
this ambiguity can be resolved by analyzing the 2D position--velocity
diagram.

Regardless of which of the above effects we include in the fit, 
the residuals 
for the outermost points ($R>0\farcs2$) still show a systematic
behaviour which indicates a velocity decrease steeper than
the expected $R^{-0.5}$ keplerian law. Unfortunately the external
points are also those with the worse SNR hence this issue cannot
be investigated further with the available data, but a possible explanation 
might be found in a slight warping of the disk at large
radii. Such warping, 
if present, does not affect the estimate of the central mass.

\begin{figure}
\centerline{\psfig{file=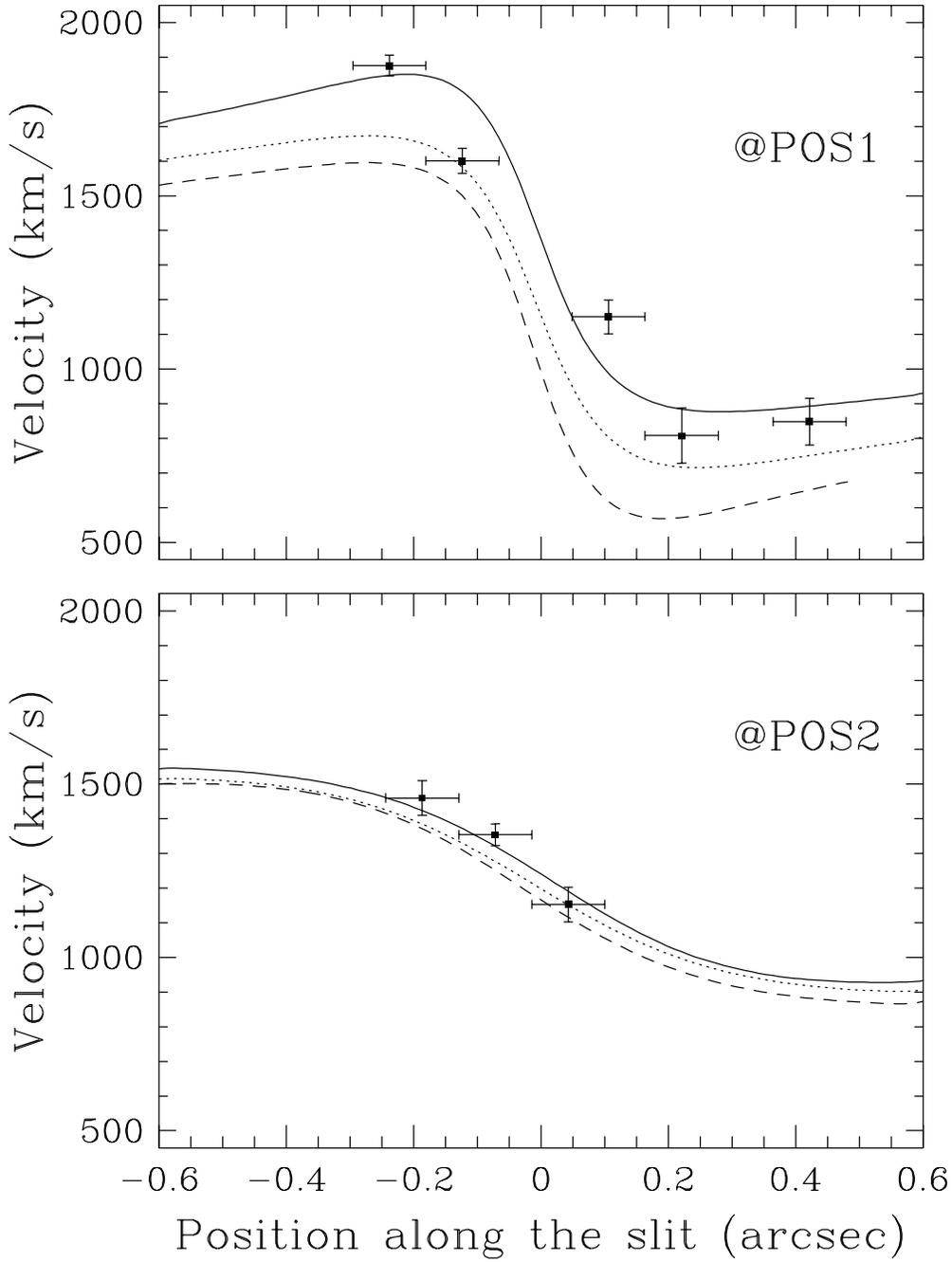,width=14cm}}
\caption{\label{fig:fitout} Predicted rotation curves at POS1 and POS2
for fit A (solid), B (dotted) and C (dashed).
Given the uncertainties on the zero--points described in the text,
the data points have been shifted in velocity
and space to match model A.}
\end{figure}

The comparison between the predictions of models A,B and C
and the velocities observed at POS1 and POS2 are shown in
Fig. \ref{fig:fitout}. Because of the uncertainty in the zero--points
of both velocity and position along the slit, which we
described in Sec. 3,
the off--nuclear data do not provide as good a constraint on the
models as one might at first expect.
This comparison shows that all three models lead to velocity gradients
compatible with the data, though as presented,
the data have been arbitrarily shifted to match model A.
If it were not for the zero--point uncertainty
the POS1 and POS2 data would allow us to unambiguously choose
between the three models.

Taking into account all possible fits which are compatible with the
data, the estimated value for the 
mass is $M_{BH} (\sin i)^2=2.0^{+0.5}_{-0.4}\xten{9}$\Mo
and $M_{BH} = (3.2\pm 0.9)\xten{9}$\Mo with $i=51^\circ$ and
its allowed range of variation.

\section{\label{sec:profiles} Analysis of the line profiles}

The line profiles are given by:
\begin{equation}
\Phi(v; S) =
\frac{\int_{S-\Delta S}^{S+\Delta S} ds \int_{B-h}^{B+h} db
\int\int_{-\infty}^{+\infty} db^\prime ds^\prime
\phi(v-V(s^\prime,b^\prime)) I(s^\prime,b^\prime) P(s^\prime-s, b^\prime-b) }
{ \int_{S-\Delta S}^{S+\Delta S} ds \int_{B-h}^{B+h} db
\int\int_{-\infty}^{+\infty} db^\prime ds^\prime
		    I(s^\prime,b^\prime) P(s^\prime-s, b^\prime-b) }
\end{equation}
where the symbols used are the same than those in the preceding equations and
$\phi(v-V)$ is the intrinsic line profile. If the motions are purely
keplerian and turbulence is negligible or less than the instrumental
FWHM this is simply a gaussian with a FWHM=430 km \isec.

Rather than attempting to carry out a full model fit to the line profiles,
which would require us to know the true surface brightness distribution
of the line within the unresolved core, we proceeded by computing
the expected line profiles using
both an exponential and a power law dependence on radius
($\Sigma(R)\propto\exp(-(R/0\farcs1))$ and
$\Sigma(R)\propto R^{-2}$ respectively).
The scaling parameters were chosen to be consistent
with the observed luminosity profile along the slit.
As noted at the end of Sec. \ref{sec:results}, the existence of a double 
peak in the observed \oii\ position--velocity diagram of 
might imply the presence of a central hole in the line
emission. To take this into account,
$\Sigma(R)$ is then multiplied by a ``hole function'' which 
forces to zero intensity all the points with R less 
than the radius of the hole.
To reproduce the observed double peaked structure the radius of the hole
must be larger than $\simeq0\farcs03$ and, moreover,
models with smaller radii predict line widths broader than those
observed. Models with hole radii larger than 0\farcs05 were discarded
since the predicted line widths at the center are much smaller
than observed. Since the hole in the emissivity distribution has a smaller
radius than the PSF, the central dark mass condensation might
be either point-like or distributed within the hole.

\begin{figure}
\centerline{\psfig{file=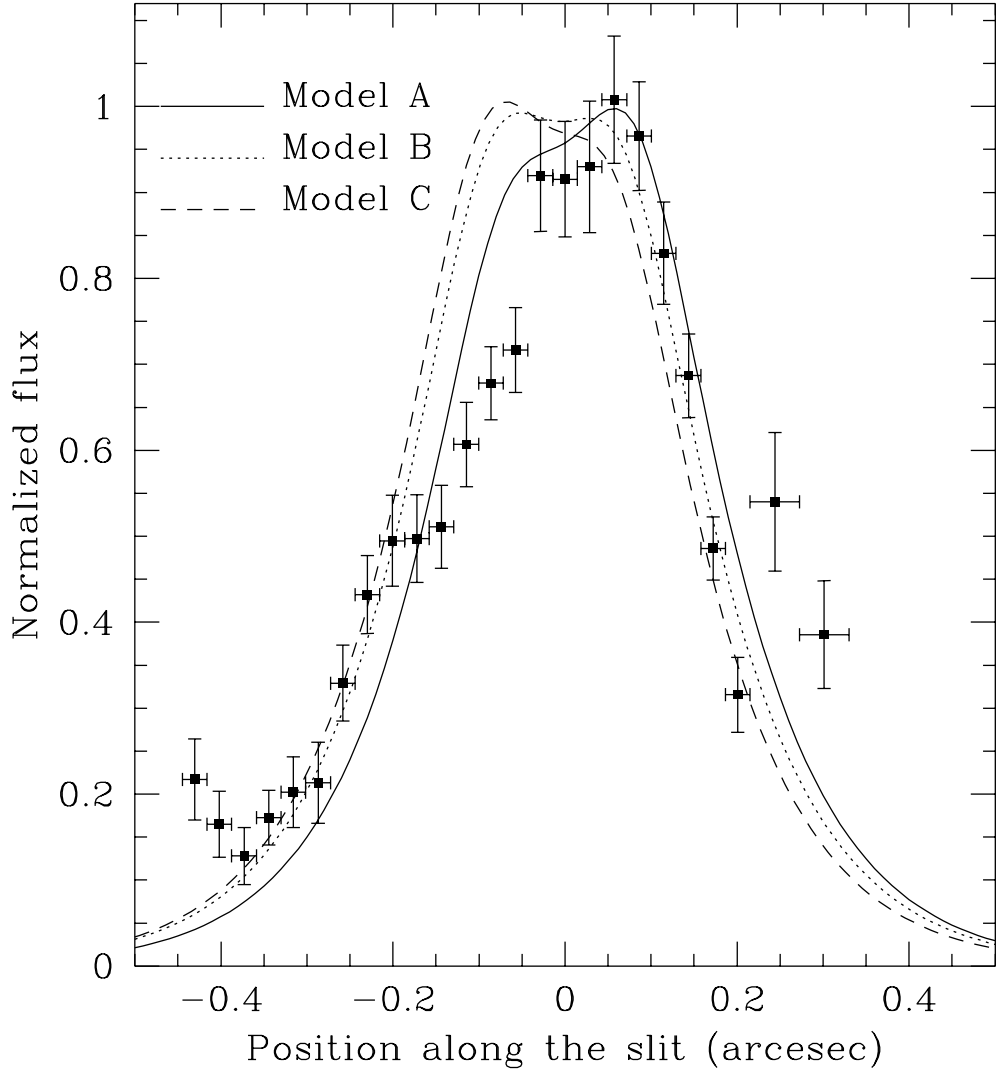,width=15cm}}
\caption{\label{fig:lumslit} Observed flux distribution
of the \oii\ line along the slit (dots with error--bars)
compared with the expected values from the exponential luminosity
distribution described in the text and the parameters derived from
model A, B and C.}
\end{figure}

The model luminosity profiles of the line along the slit 
derived for the same three representative models are compared with 
the observed \oii\ light profile in Fig. \ref{fig:lumslit}, and are
all compatible with it.
As shown above, the presence of the ``hole'' in the
emission does not significantly alter the
rotation curve, as shown above, but produces changes in the line profiles.

\begin{figure}
\centerline{\psfig{file=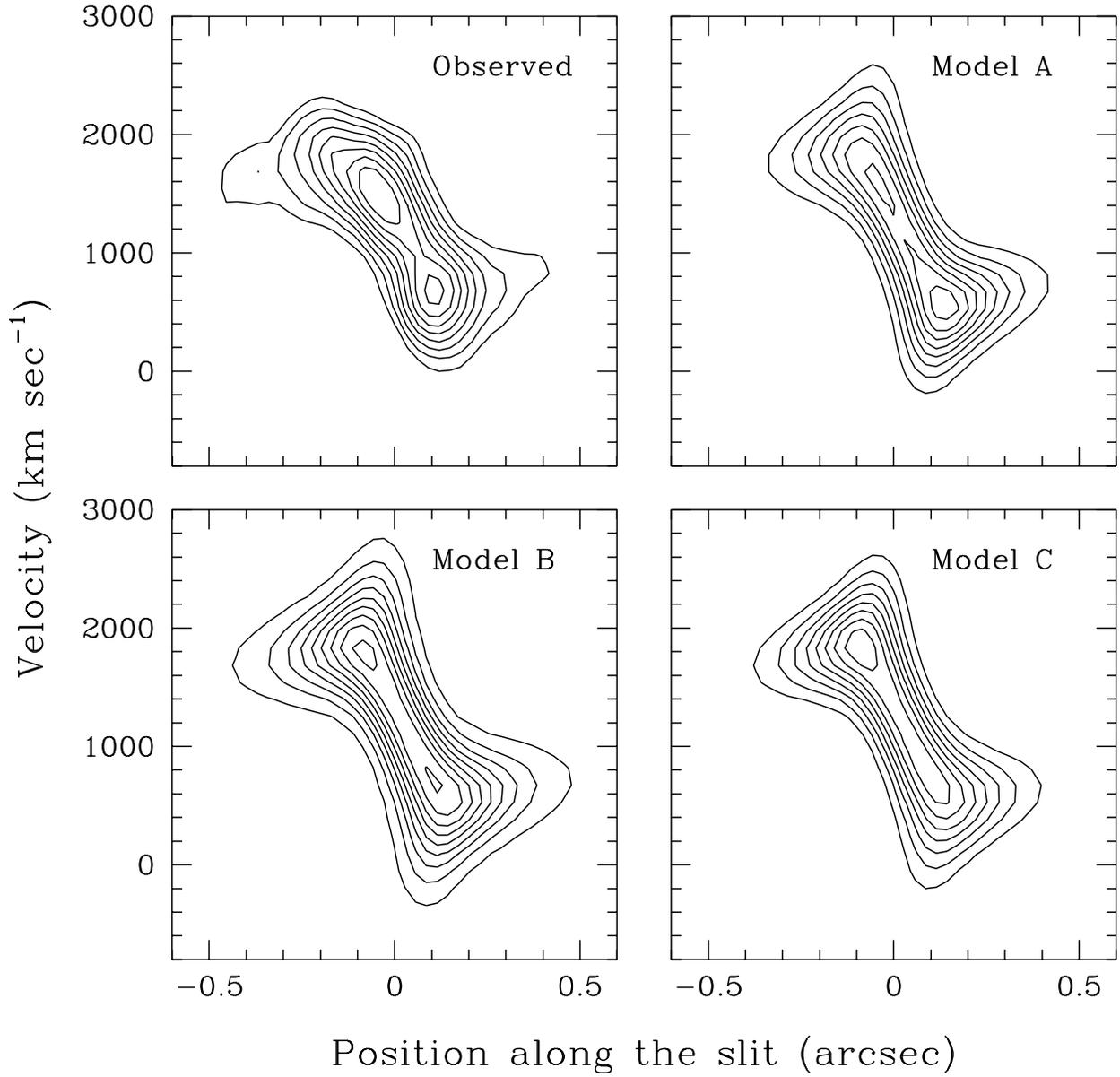,width=17cm}}
\caption{\label{fig:modcont} Predicted Velocity--space contours using
the parameters of model A, B and C and the exponential
luminosity distribution described in the text.
All levels range from 20\% to 90\% of the maximum and
are spaced by a 10\% interval.}
\end{figure}

In Fig. \ref{fig:modcont} we compare the observed and model \oii\ intensity
contours derived using the parameters from fits A, B and C, the
exponential luminosity distribution and a hole radius of 0\farcs05.
A is the model which best agrees with the data.
Models B and C, with
$\theta\simeq 0^\circ$ and $\theta> 0^\circ$ respectively ,
do not reproduce the observed position of the emission peaks.
Thus model A is the most satisfactory of the three test cases.

From the computed 2D position--velocity diagrams we can infer that
i) the choice of the intensity distribution, as long as it is
radially symmetric does not significantly alter the results;
ii) the presence of two peaks is indeed the result of a hole
in the luminosity profile;
iii) the two peaks are shifted 
with respect to the center of rotation if
the inclination angle $\theta$ is different from zero;
iv) the shift is in the direction of the observations only if
$b$ and $\theta$ have opposite signs and, since
$b>0$ as shown earlier, $\theta$ must be negative;
v) the presence of the hole
is also required to prevent the line widths in the center to
be broader than those observed.

\begin{figure}
\centerline{\psfig{file=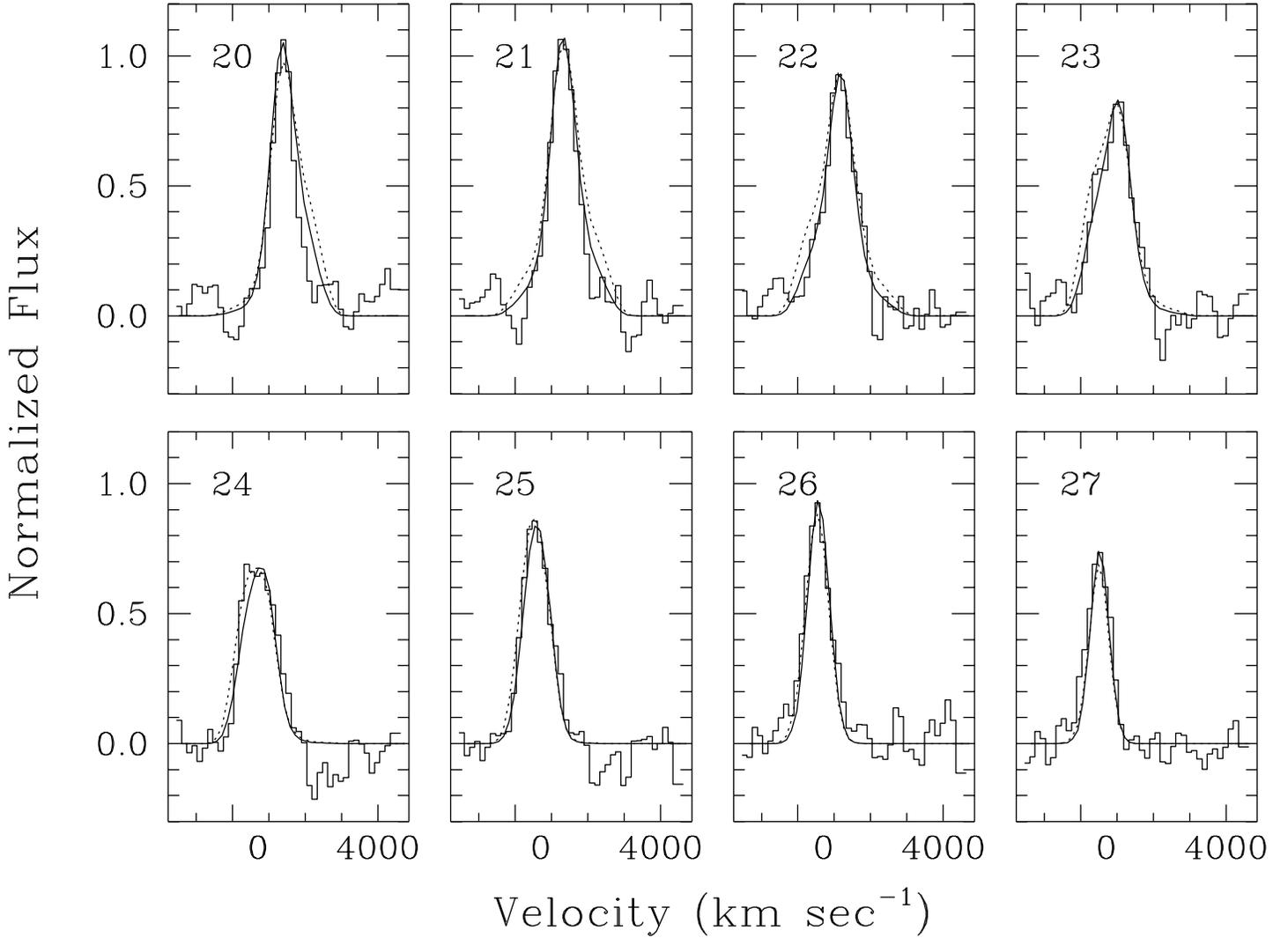,height=15cm,angle=-90}}
\caption{\label{fig:modprof} Observed \oii\ profiles (the number in the 
upper left corner represent the pixel along the slit where the profile
was extracted) compared with those predicted with the exponential 
luminosity (solid line) and the power law luminosity distributions
and the parameters of model A. To allow
us to better compare the line profiles,
the model profiles have been renormalized to those observed and shifted.}
\end{figure}

In Fig. \ref{fig:modprof} we plot the predicted line profiles compared
with those observed in the central pixels.
The solid and dotted lines are the profiles derived with
the exponential and power law luminosity distributions
respectively.
The model profiles have been scaled and re--grided to match the
pixelation of the actual data.
The agreement is remarkable especially since  this is not 
a direct fit to the profiles.
The keplerian model fully reproduces
the observed line widths and the different choices
of the luminosity distributions do not alter
this result.
Furthermore the model naturally accounts
for the shift between the position at which the FWHM is maximum and 
the point of minimum distance from the nucleus, i.e. the
peak of the continuum distribution.
This is simply a consequence of the non--null
impact parameter and angle between the slit and the line
of nodes.

In summary 
a thin--disk in keplerian motion around a central
black hole explains all the observed characteristics.
This fact strengthens 
the reliability of the derived value for the BH mass of
$M_{BH} = (3.2\pm 0.9)\xten{9}$\Mo.
The emission of \oii\  is absent in the regions
closest to the black hole ($R<3.5$pc). 
Physically this might be due to either the
gas being fully ionized or to the gas having been blown away
by the interaction with the jet.

\section{Can the mass be distributed?}

In the above sections we have demonstrated that 
$(3.2\pm 0.9)\xten{9}$\Mo\ are required to explain
the observed rotation curve and, so far, we have assumed that this mass
is point-like.

\begin{figure}
\centerline{\psfig{file=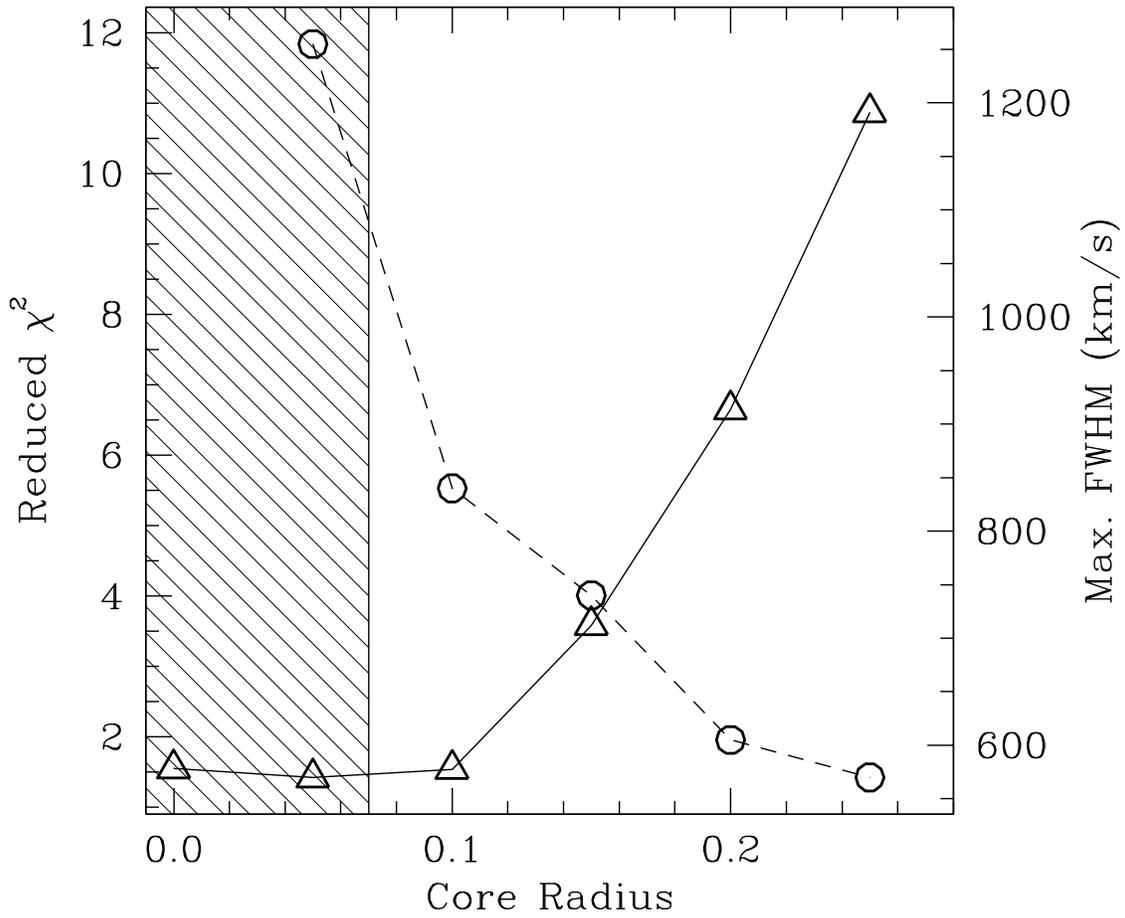,width=16cm}}
\caption{\label{fig:plummer} Solid line: value of the
minimum $\chi^2$ from the fit of the NUC rotation curve
for an assumed core radius of the Plummer model. Dashed line: 
maximum FWHM of the line profiles which can be obtained with
the Plummer model as a function of the core radius.
The shaded region indicates the range in core radii
which reproduce the observed FWHMs.}
\end{figure}

To investigate if more extended mass distributions are consistent with the 
data we have fitted the rotation curves derived with a Plummer Potential 
(e.g. \cite{binney}) with increasing core radii.
Fitting the NUC data with a core radius larger than 0\farcs05 and   
keeping all the other parameters free leads to solutions which tend to make the
impact parameter 0. Such fits are not consistent with the observations for two
reasons: i) the impact parameter has a value of 0\farcs07 as discussed
in Sec. \ref{sec:impact}, ii) decreasing the impact parameter of
the slit at the NUC position increases that of the slit at POS1
hence the models are not able to reproduce the spatial structure of the velocity field
even within the scope of our limited off-nuclear data.
Consequently we fit the data by fixing the impact parameter in
the range 0\farcs06-0\farcs08.
The minimum $\chi^2$ which can be obtained increases with increasing
core radius. Moreover to reproduce the observed rotation curve
at NUC the total mass increases.
In Fig. \ref{fig:plummer} we plot the minimum $\chi^2$ as a function
of the core radius (solid line) for those fits which
reproduce the velocity field at POS1 and POS2 and whose 
total mass is consistent with the limit of \ten{10}\Mo\ implied by the
large scale stellar dynamical measurements (\cite{vandermarel}).

Acceptable fits to the rotation curve can be found provided the core radius is
less than $0\farcs13$. However the observed radial variation of the line FWHM 
provides a more stringent constraint. The dashed line in Fig. \ref{fig:plummer}
represents the maximum FWHM of the lines which can be expected for a given core radius
(assuming an exponential luminosity distribution) and the shaded area represents
the region which matches the observations. As can be clearly seen, we must adopt 
mass distributions with core radii smaller than 0\farcs07 to match
the observed line widths. Adding a central hole to the luminosity distribution
only compounds the problem of matching the FWHM.

Such a small core radius of course places $\simeq$ 60\% of the mass
at radii smaller than that of our PSF.
As we described in Sec. 9 the finite PSF in conjunction with a central hole in the
line emissivity distribution, even in the pure black hole model, would
allow the mass to be distributed within the PSF.

If the estimated mass were uniformly distributed in a sphere
with a 0\farcs05 radius ($\simeq3.5$pc)
the mean density would be $\simeq2\xten{7}$\Mo\,pc$^{-3}$ which
is greater than the highest value encountered in the
collapsed cores of galactic globular clusters (NGC 6256 and 6325, cf. Table II
of \cite{gc}). 

The {\it total} flux estimated in the $5\times5$ pixel$^2$
nuclear region (0\farcs28$\times$0\farcs28 $\simeq$ 16$\times$16 pc$^2$)
from F547M,F555W WFPC2 archival images
is $\simeq$ 5.3\xten{-16} erg cm$^{-2}$ \isec\ \AA$^{-1}$
which corresponds roughly to 3.2\xten{6}\Lo\ at 15Mpc in the V band.
Consequently the mass-to-light ratio in the V band is
$M/L_V\simeq$110 $\Mo/L_{V\odot}$ where $L_{V\odot}$ is the V luminosity
of the sun ($L_{V\odot}=0.113\Lo$).
Such mass-to-light ratio is uncomfortably high; indeed
from stellar population synthesis $M/L_V<20$ (e.g. \cite{bruzual}).
The above considerations suggest that the mass
condensation in the central $R<5$pc of the nucleus of M87 cannot
be a supermassive cluster of ``normal'' evolved stars.
If it is not a supermassive black hole, it must nonetheless be
quite an ``exotic'' object such as a massive cluster of 
neutron stars or other dark objects.
A more extensive discussion of such possibilities has been given
in \cite{vandermarel97}.
We concur with their general conclusion that these alternatives
are both implausible and contrived.

\section{Summary and Conclusions}

We have presented the results of HST FOC f/48 high spatial resolution long--slit
spectroscopy of the ionized circumnuclear gas disk of M87,
at three spatially separated locations 0\farcs2 apart.

We have analyzed these data and, in particular, the \oii\ emission lines 
and derived rotation curves which extend to a distance of
$\sim$1\arcsec\ from the nucleus. Within the
uncertainties, these data are insensitive to density variations
over a broad range of values which are larger than the
constraints on density derived from the FOS archive data.

Our rotation curve is compatible with that obtained 
form the archival FOS data, within their substantially
larger intrinsic uncertainties. 
Furthermore we have verified that this applies to all 
emission lines (\hb, \oiii, \ha, \nii\ and \sii) measured with FOS which implies that 
we have not been misled by ionization conditions of the gas.

To analyze our data we have first constructed a simple
analytical model for a thin
keplerian disk around a central mass condensation, and fitted
the model function to the observed rotation curve.
Since the number of free parameters is large we carried out trial
minimization of the residual errors by using different estimates for the values
of the key parameters. This procedure allowed us to construct a series 
of self--consistent solutions as well as to highlight the
sensitivity of the final solutions to the different choices of initial
estimates for the free parameters. Using this simple model we derived two
extreme sets of self--consistent solutions which provide good fits
to the observational data.

There is marginal evidence for a warp of the disk
in the outermost ($R>0\farcs2$) points but this has little effect
on our mass estimate.

We then conducted a more realistic analysis incorporating
the finite slit width, the spatial PSF and the intrinsic luminosity
distribution of the gas.
This analysis showed that a thin keplerian disk with a central hole
in the luminosity function provides a good match to our data.
We presented
three representative models (A, B and C) which encompass the range
of variation of the line of nodes and used these to compute the line
profiles and 2D position--velocity diagrams for the \oii\ lines.
Model A best reproduces the observations, and 
the resulting parameters of the disk are
$i=51^\circ$, $\theta=-9^\circ$, $V_{sys}=1290$ km \isec\
and a corresponding mass of $(3.2\pm0.9)\xten{9}$\Mo,  
where the error in the mass allows for the uncertainty of
each of the parameter (Tab. 4).
We showed that this mass must be concentrated within a sphere
of less than 3.5 pc and concluded that the most likely
explanation is a supermassive black hole.

To make further progress there are a number of possibilities the easiest of 
which is to make a more comprehensive and higher signal-to-noise 2D 
velocity map of the disk to better constrain its parameters.  We note in 
passing that recently there has been considerable progress in modeling 
warped disks (\cite{pringle}, \cite{liviob}) and this treatment could be 
applied to such improved data to investigate the origin of the 
apparent steeper than keplerian fall off in rotation velocity beyond a 
radius of 0\farcs2 that we alluded to above.

The biggest limitation of the present data is that, even by observing with HST
at close to its optimal resolution at visible wavelengths, some of the important features
of the disk kinematics are subsumed by the central PSF. Until a larger space based 
telescope becomes available, the best we can do is to study the gas disk in Ly$\alpha$
and gain the Rayleigh advantage in resolution by moving to the UV.
This approach may run into difficulties because of geocoronal Ly$\alpha$
emission and the effects of obscuration.
Nevertheless this may be the only way to proceed because of the difficulty of
detecting the high velocity wings which characterize the stellar absorption lines
in the presence of a supermassive black hole. 

\acknowledgements

A.M. acknowledges partial support through GO grant G005.44800 from
Space Telescope Science Institute, which is operated by the Association
of Universities for Research in Astronomy, Inc., under NASA contract
NAS 5--26555.

A.C. acknowledges support from the STScI visitor program.

We thank E. Oliva for kindly providing his
compilation of atomic parameters and his code to derive line
emissivities.
We thank Stefano Casertano Massimo Stiavelli and Roeland van der Marel
for stimulating discussions and suggestions which improved the
analysis and Mario Livio for
a careful reading of the manuscript.
We thank Robert Jedrzejewski, Mark Voit and Dorothy 
Fraquelli for their assistance
during the observations.
We thank the anonymous referee and the scientif editor, Dr. Greg Bothun,
for useful comments and suggestion which improved this paper.

\clearpage

\begin{deluxetable}{lccccl}
\tablenum{1}
\tablecaption{Log of observations.}
\tablewidth{7in} \tablehead{
\colhead{Target} & \colhead{Dataset}
& \colhead{Date} & \colhead{Int. Time (s)} &\colhead{Format} &\colhead{Description}  }
\startdata
M87	 & X3E40101T	& Jul. 25, 96	& 297	& 1024x512   & interactive acq. \nl
M87	 & X3E40102T	& Jul. 25, 96	& 2169	& 1024x512   & spectrum @POS1\nl
M87	 & X3E40103T	& Jul. 25, 96	& 600	& 1024x512   & internal flat \nl
M87	 & X3E40104T	& Jul. 25, 96	& 600	& 1024x512   & internal dark \nl
M87	 & X3E40105T	& Jul. 25, 96	& 2169	& 1024x512   & spectrum \#1 @NUC\nl
M87	 & X3E40106T	& Jul. 25, 96	& 600	& 1024x512   & internal flat \nl
M87	 & X3E40107T	& Jul. 25, 96	& 2169	& 1024x512   & spectrum @POS2\nl
M87	 & X3E40108T	& Jul. 25, 96	& 600	& 1024x512   & internal flat \nl
M87	 & X3E40109T	& Jul. 25, 96	& 2597	& 1024x512   & spectrum \#2 @NUC\nl
M87	 & X3E4010AT	& Jul. 25, 96	& 600	& 1024x512   & internal flat \nl
M87	 & X3E4010BT	& Jul. 25, 96	& 2597	& 1024x512   & spectrum \#3 @NUC\nl
M87	 & X3E4010CT	& Jul. 25, 96	& 600	& 1024x512   & internal flat \nl
M87	 & X3E4010DT	& Jul. 25, 96	& 2597	& 1024x512   & spectrum \#4 @NUC\nl
 & & & & & \nl
47 Tuc	 & X34I0108T	& Apr. 4, 96	& 477	& 1024x256z  & spectrum      \nl 
47 Tuc	 & X34I0109T	& Apr. 4, 96	& 600	& 1024x256z  & internal flat \nl 
NGC 6543 & X3BD0102T	& Sep. 10, 96	& 682	& 1024x512   & spectrum      \nl
NGC 6543 & X3BD0105T	& Sep. 10, 96	& 500	& 1024x512   & internal flat \nl
\enddata
\end{deluxetable}

\begin{deluxetable}{lllll}
\tablenum{2}
\tablecaption{Archival FOS data.}
\tablewidth{0pt} \tablehead{
\colhead{Target} & \colhead{Datasets}	& \colhead{Aperture(\arcsec)}
& \colhead{R(\arcsec)} & \colhead{PA($^\circ$)} }
\startdata
POS1	& Y2760104T		& 0\farcs26	& 0.35    & 135 \nl
POS2	& Y2760107T,Y2760108T	& 0\farcs26	& 0.56    & 153  \nl
POS4	& Y2D90105T		& 0\farcs26	& 0      &  0 \nl
POS4b	& Y2KZ0104T		& 0\farcs26	& 0\tablenotemark{*}     & 0\tablenotemark{*} \nl
POS5	& Y2D90106T,Y2D90107T	& 0\farcs26	& 0.25   &  21 \nl
POS6	& Y2D90108T,Y2D90109T	& 0\farcs26	& 0.25   &  201 \nl
POS7	& Y2KZ0105T		& 0\farcs26	& 0.25    & 291 \nl
POS8	& Y2KZ0106T		& 0\farcs26	& 0.25    & 111 \nl
POS9a	& Y2KZ0205T,Y2KZ0206T	& 0\farcs09	& 0.086   & 21 \nl
POS9b	& Y2kZ0309T,Y2KZ030AT	& 0\farcs09	& 0.086\tablenotemark{*}   & 21\tablenotemark{*} \nl
POS10a	& Y2KZ0207T,Y2KZ0208T	& 0\farcs09	& 0       & 0 \nl
POS10b	& Y2KZ0307T,Y2KZ0308T	& 0\farcs09	& 0\tablenotemark{*}  & 0\tablenotemark{*} \nl
POS11a	& Y2KZ0209T,Y2KZ020AP	& 0\farcs09	& 0.086   & 201 \nl
POS11b	& Y2KZ0305T,Y2KZ0306T	& 0\farcs09	& 0.086\tablenotemark{*}   & 201\tablenotemark{*} \nl
\tablenotetext{*}{Positions are nominally the same as the preceding ones
but there was a small misplacement during observations.}
\enddata
\end{deluxetable}

\begin{deluxetable}{lcccccc}
\tablenum{3}
\tablecaption{ Velocities from archival FOS observations.}
\tablewidth{0pt} \tablehead{
\colhead{Line} & \colhead{POS1} & \colhead{POS2} & \colhead{POS4} &
\colhead{POS4b}& \colhead{POS5} & \colhead{POS6} }
\startdata
\ \hb\wl 4861.3		&1073 & 1019 & 1359 & 1549 & 1811 &  863  \nl
\ \oiii\wl 4958.9	&981  & 1001 & 1402 & 1133 & 1917 &  844  \nl
\ \oiii\wl 5006.9	&991  & 1000 & 1210 & 1212 & 1900 &  766    \nl
\ \oi\wl 6300.3		&1193 & 1003 & \nodata  & \nodata & 1795 &  852    \nl
\ \nii\wl 6548.0	&1093 & 978  & \nodata   & \nodata   & 1778 &  820  \nl
\ \ha\wl 6562.8		&1166 & 972  & \nodata   & \nodata   & 1795 &  831  \nl
\ \nii\wl 6584.0	&1053 & 950  & \nodata   & \nodata   & 1809 &  846  \nl
\ \sii\wl 6716.4	&1073 & 1036 & \nodata  & \nodata   & 1772 &  920   \nl
\ \sii\wl 6730.8	&1100 & 1066 & \nodata  & \nodata   & 1827 &  950   \nl
\tablevspace{5pt}
                  & 1080$\pm$70 & 1005$\pm$35 & 1323$\pm$100 & 1298$\pm$200 &
		  1820$\pm$50 & 854$\pm$50 \nl
\tablevspace{10pt} \hline\hline \tablevspace{3pt}
\multicolumn{1}{c}{Line} & POS 7 & POS 8 & POS9a & POS9b & POS11a & POS11b  \nl
\tablevspace{3pt} \hline \tablevspace{10pt}
\ \hb\wl 4861.3		& 1060& 1418 & 1656 & 1450    & 868  & 679   \nl
\ \oiii\wl 4958.9	& 800& 1120 & 1649 & 1763    & \nodata & 326   \nl
\ \oiii\wl 5006.9	& 850& 1259 & 1615 & 1651   & 752  & 499   \nl
\ \oi\wl 6300.3		&\nodata  & \nodata  & 1649 & \nodata    & \nodata   & \nodata   \nl
\ \nii\wl 6548.0	&\nodata  & \nodata  & 1682 & \nodata       & \nodata & \nodata  \nl
\ \ha\wl 6562.8		&\nodata  & \nodata  & 1716 & \nodata       & \nodata & \nodata   \nl
\ \nii\wl 6584.0	&\nodata  & \nodata  & 1639 & \nodata       & \nodata & \nodata   \nl
\ \sii\wl 6716.4	&\nodata  & \nodata  & 1675 & \nodata       & \nodata   & \nodata    \nl
\ \sii\wl 6730.8	&\nodata  & \nodata  & 1698 & \nodata      & \nodata   & \nodata    \nl
\tablevspace{5pt}
\ &900$\pm$130 & 1265$\pm$160 & 1660$\pm$35 & 1621$\pm$160 & 810$\pm$60 &
 520$\pm$150   \nl
\enddata
\end{deluxetable}

\begin{deluxetable}{ccc}
\tablenum{4}
\tablecaption{Ranges of variation of the parameters of
the fit.}
\tablewidth{0pt} \tablehead{
\colhead{Parameter} & \multicolumn{2}{c}{Without PSF ($\chi^2<2.5)$} }
\startdata
                   & $47^\circ<i<55^\circ$ & $55^\circ<i<65^\circ$ \nl
$M_{BH}(\sin i)^2\,\tablenotemark{a}$ &  1.64--1.83 & 1.65--1.91 \nl
$b$                & 0\farcs07--0\farcs085 & 0\farcs058--0\farcs074 \nl
$\theta$           & -5$^\circ$--3$^\circ$ & -5$^\circ$--4$^\circ$ \nl
$V_{sys}\tablenotemark{b}$          & 1175--1260 & 1150--1280 \nl
$cpix$             & 22.6--22.9 & 22.6--22.8 \nl
\tablevspace{10pt}
\hline\hline \tablevspace{3pt}
Parameter	   & \multicolumn{2}{c}{With PSF ($\chi^2<2.5$)}  \nl
\tablevspace{3pt}
\hline
\tablevspace{3pt}
                   & $39^\circ<i<55^\circ$ & $55^\circ<i<65^\circ$ \nl
$M_{BH}(\sin i)^2\,\tablenotemark{a}$ & 1.65--2.31 & 1.94--2.48 \nl
$b$                & 0\farcs076--0\farcs085 & 0\farcs064--0\farcs085 \nl
$\theta$           & -11$^\circ$--13$^\circ$ & -15$^\circ$--11$^\circ$ \nl
$V_{sys}\tablenotemark{b}$          & 1085--1300 & 1080--1355 \nl
$cpix$             & 22.4--23.1 & 22.5--23.1 \nl
\tablenotetext{a}{in units of \ten{9}\Mo.}
\tablenotetext{b}{ km \isec.}
\enddata
\end{deluxetable}

\clearpage

\clearpage

\clearpage

\end{document}